\documentclass[twocolumn]{aastex61}
\usepackage[maxfloats=40]{morefloats} 
\usepackage{color}
\usepackage{amsmath}
\usepackage{catoptions}
\usepackage[nameinlink]{cleveref}
\usepackage[figure,figure*]{hypcap}

\newcommand{\Rom}[1]{\uppercase\expandafter{\romannumeral #1\relax}}
\creflabelformat{equation}{#2\textup{#1}#3}

\begin{document}
\submitjournal{\apj}
\accepted{November 25, 2019}
\title{Multiwavelength Polarimetry of the Filamentary Cloud IC5146: \Rom{2}. Magnetic Field Structures}

\author[0000-0002-6668-974X]{Jia-Wei Wang}
\affiliation{Institute of Astronomy and Department of Physics, National Tsing Hua University, Hsinchu 30013, Taiwan}
\affiliation{Academia Sinica Institute of Astronomy and Astrophysics, P.O. Box 23-141, Taipei 10617, Taiwan}
\email{jwwang@gapp.nthu.edu.tw; slai@phys.nthu.edu.tw}

\author[0000-0001-5522-486X]{Shih-Ping Lai}
\affiliation{Institute of Astronomy and Department of Physics, National Tsing Hua University, Hsinchu 30013, Taiwan}

\author[0000-0002-9947-4956]{Dan P. Clemens}
\affiliation{Institute for Astrophysical Research, Boston University, 725 Commonwealth Avenue, Boston, MA 02215, USA}

\author[0000-0003-2777-5861]{Patrick M. Koch}
\affiliation{Academia Sinica Institute of Astronomy and Astrophysics, P.O. Box 23-141, Taipei 10617, Taiwan}

\author[0000-0003-4761-6139]{Chakali Eswaraiah} 
\affiliation{CAS Key Laboratory of FAST, NAOC, Chinese Academy of Sciences, Beijing 100101, People’s Republic of China}
\affiliation{National Astronomical Observatories, Chinese Academy of Sciences, Datun Road, Chaoyang District, Beijing 100101, People’s Republic of China}

\author[0000-0003-0262-272X]{Wen-Ping Chen}
\affiliation{Institute of Astronomy, National Central University, Chung-Li 32054, Taiwan}

\author{Anil K. Pandey} 
\affiliation{Aryabhatta Research Institute of Observational-Sciences (ARIES), Nainital - 263001, India}

\begin{abstract}
The IC5146 cloud is a nearby star-forming region in Cygnus, consisting of molecular gas filaments in a variety of evolutionary stages. We used optical and near-infrared polarization data toward the IC5146 cloud, reported in the first paper of this series, to reveal the magnetic fields in this cloud. Using the newly released $Gaia$ data, we found that the IC5146 cloud may contain two separate clouds: a first cloud, including the densest main filament at a distance of $\sim$600 pc, and a second cloud, associated with the Cocoon Nebula at a distance of $\sim$800 pc. The spatially averaged H-band polarization map revealed a well-ordered magnetic field morphology, with the polarization segments perpendicular to the main filament but parallel to the nearby sub-filaments, consistent with models assuming that the magnetic field is regulating cloud evolution. We estimated the magnetic field strength using the Davis-Chandrasekhar-Fermi method, and found that the magnetic field strength scales with volume density with a power-law index of $\sim0.5$ in the density range from $N_{H_2}\sim$ 10 to 3000 cm$^{-3}$, which indicates an anisotropic cloud contraction with a preferred direction along the magnetic field. In addition, the mass-to-flux ratio of the cloud gradually changes from subcritical to supercritical from the cloud envelope to the deep regions. These features are consistent with strong magnetic field star-formation models and suggest that the magnetic field is important in regulating the evolution of the IC5146 cloud.

\end{abstract}

\keywords{ISM: clouds --- ISM: magnetic fields --- ISM: structure --- ISM: individual objects (IC5146) --- Polarization}

\section{INTRODUCTION}
\subsection{Magnetic Fields in Filamentary Clouds}
\label{sec:intro}
Observations in the past few decades revealed that dense cloud cores predominately appear in clusters and form from magnetized and turbulent molecular clouds \citep{mk07}. These molecular clouds are often elongated or filamentary over parsec scales, and the structures are believed to directly influence the star-formation process \citep{an14}. However, how prestellar cores form in these clouds remains poorly understood. Theoretical studies suggest that both turbulence and magnetic fields inside the clouds may control the star formation and cloud evolution, but their relative importance is still under debate (as reviewed in \citealt[]{mk07}). Some numerical simulations suggest that increasing the magnetic field strength would decrease the predicted, overly high, star-formation rate to values close to the observed rate \citep{na08,pr08}. In addition, the magnetic field can also guide the collapse of clouds \citep[e.g.,][]{na08,va14} and stabilize the filamentary structures \citep[e.g.,][]{in15,se15}. In contrast, simulations assuming weak magnetic fields find that compression and fragmentation from supersonic turbulence can reproduce core mass functions, consistent with the observed stellar initial mass functions \citep{pa02,ma04}. Due to the difficulty of measuring the magnetic field structures of molecular clouds, from large to small scales, past observations have been insufficient to settle this debate.

The $Herschel$ Gould Belt Survey \citep{an10} showed that filamentary structures are ubiquitous in both quiescent and active star-forming regions. Bound prestellar cores and deeply embedded protostars were primarily found in the filaments with column densities $N_{H_2}> 7\times10^{21}$ cm$^{-2}$ \citep{an10,mo10}. In addition, the observed filamentary structures seem to share a universal characteristic width of $\sim$ 0.1 pc, regardless of their central column density or environment \citep{ar11,an14}. These results favor a scenario in which the filaments are first generated in molecular clouds, due to large-scale magneto-hydrodynamic (MHD) turbulence, and fragment into prestellar cores due to gravitational instability \citep{me10,mi10,wa10}.

In contrast, optical and infrared polarization observations using reddened background starlight show that the orientations of magnetic field are commonly perpendicular to the long axis of main filaments (the parsec-scale, overall filamentary structure) \citep[e.g.,][]{fr10,li13}. Recent $Planck$ survey data \citep{pl16} further show that most filaments tend to be parallel to magnetic fields in the diffuse interstellar medium, but become perpendicular to magnetic fields in dense molecular clouds. These results favor strong magnetic field models in which the turbulent compression and global gravitational contraction in clouds are regulated by the magnetic field \citep[e.g.,][]{na08,va14,in15}. 

In addition to the magnetic field morphology, measurements of magnetic field strength are essential to evaluate the relative importance among gravity, magnetic fields, and turbulence. Weak magnetic field models assume an environment where magnetic fields are too weak to regulate the isotropic gravitational collapse and turbulent compression, hence magnetic fields are expected to efficiently scale with volume density by a power-law index of 2/3 via gas contraction \citep{me66}. On the other hand, strong magnetic field models describe an environment where magnetic fields are sufficiently strong to impede the gravitational collapse and turbulent compression, and so clouds tend to collapse/contract along magnetic fields, resulting in magnetic field strengths scaling with volume density by a shallower power-law index of $\sim0.5$ \citep[e.g.,][]{mo91}. Furthermore, the two theories have very different predictions for the mass-to-flux ratio ($\lambda$), the relative strength of the gravitational potential compared to the magnetic field flux. Weak magnetic field models predict a broad range of $\lambda$ from 1 to 10, while strong magnetic field models expect $\lambda\approx1$ \citep{cr12}. As a result, for clouds with comparable densities, the probability density function (PDF) of magnetic field strength is expected to be broad and near-uniform in weak magnetic field environments, but narrow and Gaussian-like for strong magnetic field environments \citep{cr10}.

\citet{cr10} summarized Zeeman effect measurements, the most direct method to measure the magnetic field strength, toward 137 H\Rom{1} and molecular clouds, and found that the measured magnetic field strengths scaled with volume density with a power index of 0.65. Their Bayesian analysis favored a uniform PDF of total field strengths over a Gaussian-like PDF. Both of these features suggested that magnetic fields are often too weak to dominate the star formation process. However, these Zeeman measurement samples were taken from numerous different types of clouds that do not have evolutionary connections \citep{li14,li15}, and also most of the Zeeman measurements only covered a small area of each cloud. Therefore, using these statistics to infer cloud evolution in weak magnetic field condition is questionable, and observations that resolve magnetic field strengths within a single cloud (e.g., \citet{ma12,ho17}) are thus essential for elucidating cloud evolution.

\subsection{The IC5146 Cloud}
The IC5146 cloud is a cloud system considered to be forming from converging flows due to large-scale MHD turbulence \citep{ar11}, and hence was chosen as our target to study the role of magnetic fields in a turbulent environment. This cloud is a nearby star-forming region in Cygnus, composed of an H\Rom{2} region (the Cocoon Nebula) and a dark cloud extending to the west. $Herschel$ observations reveal that the dark cloud consists of a long, major, filamentary structure and several sub-filament structures extended from or within the main filament \citep{ar11}. The Cocoon Nebula and the main filament are both known active star-forming regions, although the YSO populations indicate that these two star-forming regions are likely in different evolutionary stages \citep{ha08,du15}. On the other hand, a part of the dark clouds still remains in quiescent stage \citep{ar11}. The variety of filamentary features in the IC5146 system suggests that it is an ideal target for investigating the formation and evolution of these filaments \citep{jo17}.

In the first paper of this series, \citet{wa17} (hereafter Paper \Rom{1}) reported our measurements of starlight polarization across this cloud at both optical and near-infrared wavelengths. The analysis of polarization efficiency suggested that the dust grains are likely still aligned with the magnetic fields within the IC5146 cloud for $A_V$ up to at least 20 mag, and thus our polarization data are capable of tracing the magnetic fields in and around the cloud, instead of simply within the skin layer of clouds as found in previous studies \citep{go95}. \citet{wa19} further reported the JCMT 850~$\mu$m continuum polarization observations toward a sub-parsec scale hub-filament system embedded within the IC5146 main filament. The observed sub-parsec scale magnetic field is likely inherited from the parsec-scale, but was modified by the large-scale contraction of the main filament. At the same time, this sub-parsec scale magnetic field is likely important in guiding the surrounding filaments toward the dense central hub and also in shaping the star-forming clumps. Understanding the large-scale evolution of filaments and magnetic fields is crucial to further probe the complete view of star-formation within a cloud.

Previously published distance estimates for the IC5146 cloud have been inconsistent. \citet{ha08} derived a distance of 950 pc based on a comparison of the absolute magnitudes of the B-type stars within the Cocoon Nebula with those within the Orion Nebula Cluster. They argued that the IC5146 dark cloud and the Cocoon Nebula are equidistant, because their morphology and velocity distribution seems to be connected. However, \citet{la99} estimated a distance of $460\substack{+40 \\ -60}$\, pc for the dark clouds, via comparing the number of low-extinction foreground stars to those predicted from galactic models. In Paper \Rom{1}, we selected the stars with both polarization and pre-$Gaia$ parallax distance measurements near the IC5146 cloud sky area, and showed that polarization percentage rose significantly at a distance of $\sim400$ pc, which was consistent with the \citet{la99} distance estimate. The recent $Gaia$ Data Release 2 (DR2) data \citep{gaia16,gaia18} provides the best stellar parallax measurements to date, and \citet{dz18} estimated a distance of $813\pm106$ pc based on the parallax measurements toward the embedded young stellar objects (YSOs) within the Cocoon Nebula. The origin of these inconsistent distance estimation is still unclear. The reason may be because the dark cloud and the Cocoon Nebula are not equidistant, or some of the estimation assumptions are incorrect.

In this paper, we aim to explore magnetic field properties on the parsec scale, based on our starlight polarization measurements, and investigate how the magnetic field correlates with the filamentary structure. In \autoref{sec:dist}, we reexamine the distance of the Cocoon Nebula and the associated dark clouds using the new $Gaia$ data.
In \autoref{sec:B_morphology}, we show the large-scale magnetic field morphology based on the polarization measurements toward the IC5146 cloud. \autoref{sec:B_str} presents an estimate of magnetic field strengths. \autoref{sec:Bvsn} reports the results of a Bayesian analysis on how magnetic field strength scales with density. In \autoref{sec:discussion}, we discuss the results of the analyses and their implications, and \autoref{sec:summary} presents our conclusions. In the forthcoming Paper \Rom{3}, we will present our molecular line observations toward IC5146 to investigate whether the magnetic field is sufficiently strong to regulate the gas kinematics.

\section{Analysis}\label{sec:ana}
\subsection{The Distances to the IC5146 system}\label{sec:dist}

\begin{figure}[hbtp]
\includegraphics[width=\columnwidth]{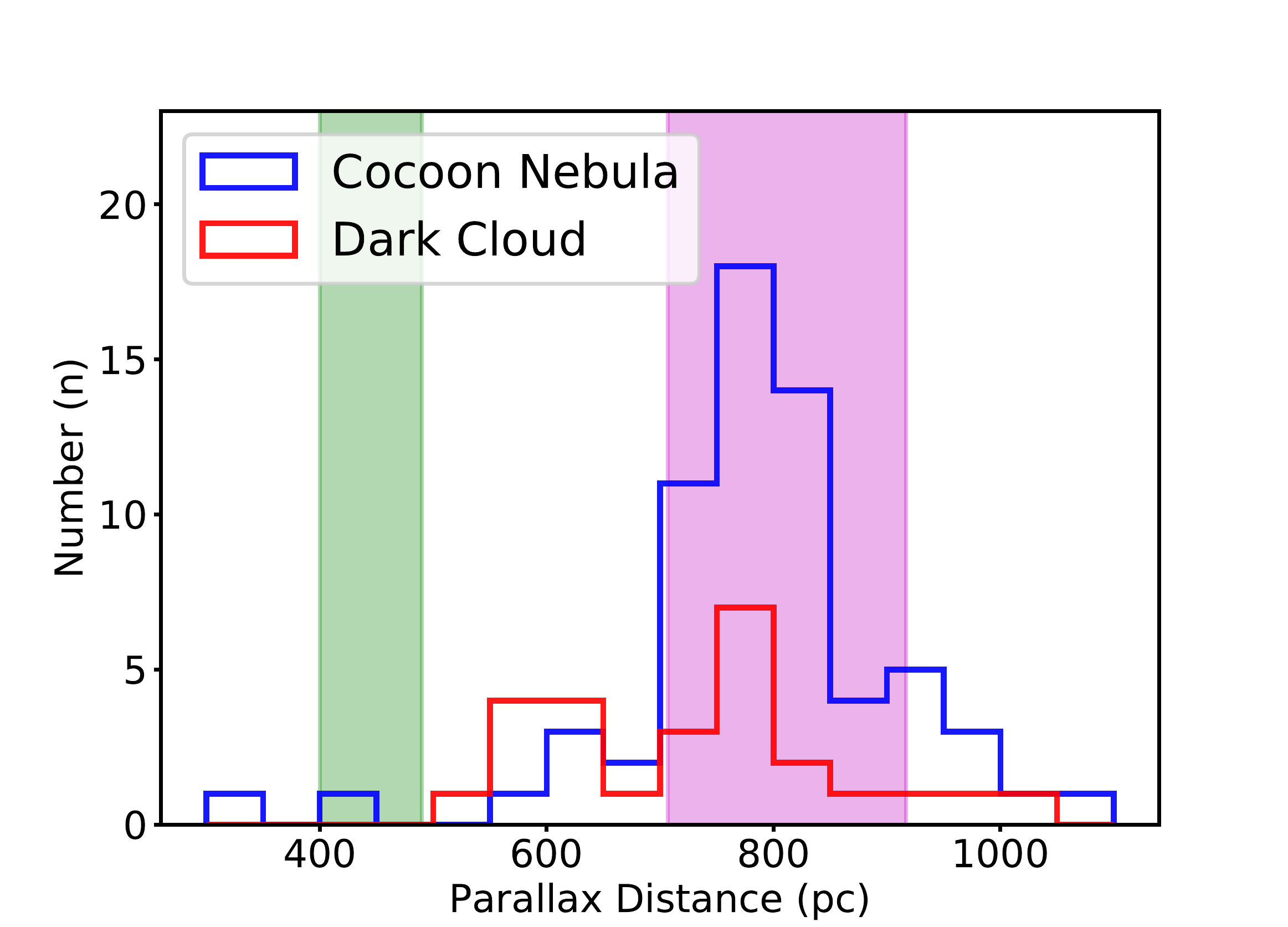}
\caption{Histogram of trigonometric parallax distances to the YSOs in the Cocoon Nebula (blue) and in the dark cloud (red), as shown in \autoref{fig:distance_map}. The YSOs were identified in \citet{ha08} using $Spitzer$ data. The parallax distances were measured by Gaia DR2, and we discard those YSOs whose trigonometric parallax measurements are less than 5$\sigma$. The green and magenta regions label the previously published distances of $460\substack{+40 \\ -60}$ pc and $813\pm106$ pc. The distances of YSOs in the Cocoon Nebula are largely in the range of $813\pm106$ pc. However, the distances of YSOs in the dark cloud show two peaks, at distances of $\sim$600 and $\sim$800 pc, suggesting that the dark cloud may be composed of multiple clouds at different distances. }\label{fig:histo_d}
\end{figure}

The key issue that causes the ambiguity of the distance estimation of the IC5146 system is whether the Cocoon Nebula and the dark clouds are equidistant. The distance estimation using the stars or YSOs within the Cocoon Nebula showed a greater distance of 800--1000 pc \citep[e.g.,][]{ha08,dz18} while the distance estimates toward the dark clouds favor a distance of 400--500 pc \citep[e.g.,][]{la99,wa17}. Hence, we seek to test whether (1) the Cocoon Nebula and the dark clouds are equidistant, but one of the distance estimation is incorrect or (2) the Cocoon Nebula and the dark clouds are not equidistant \citep{ha08}.

\begin{figure*}[hbtp]
\includegraphics[width=\textwidth]{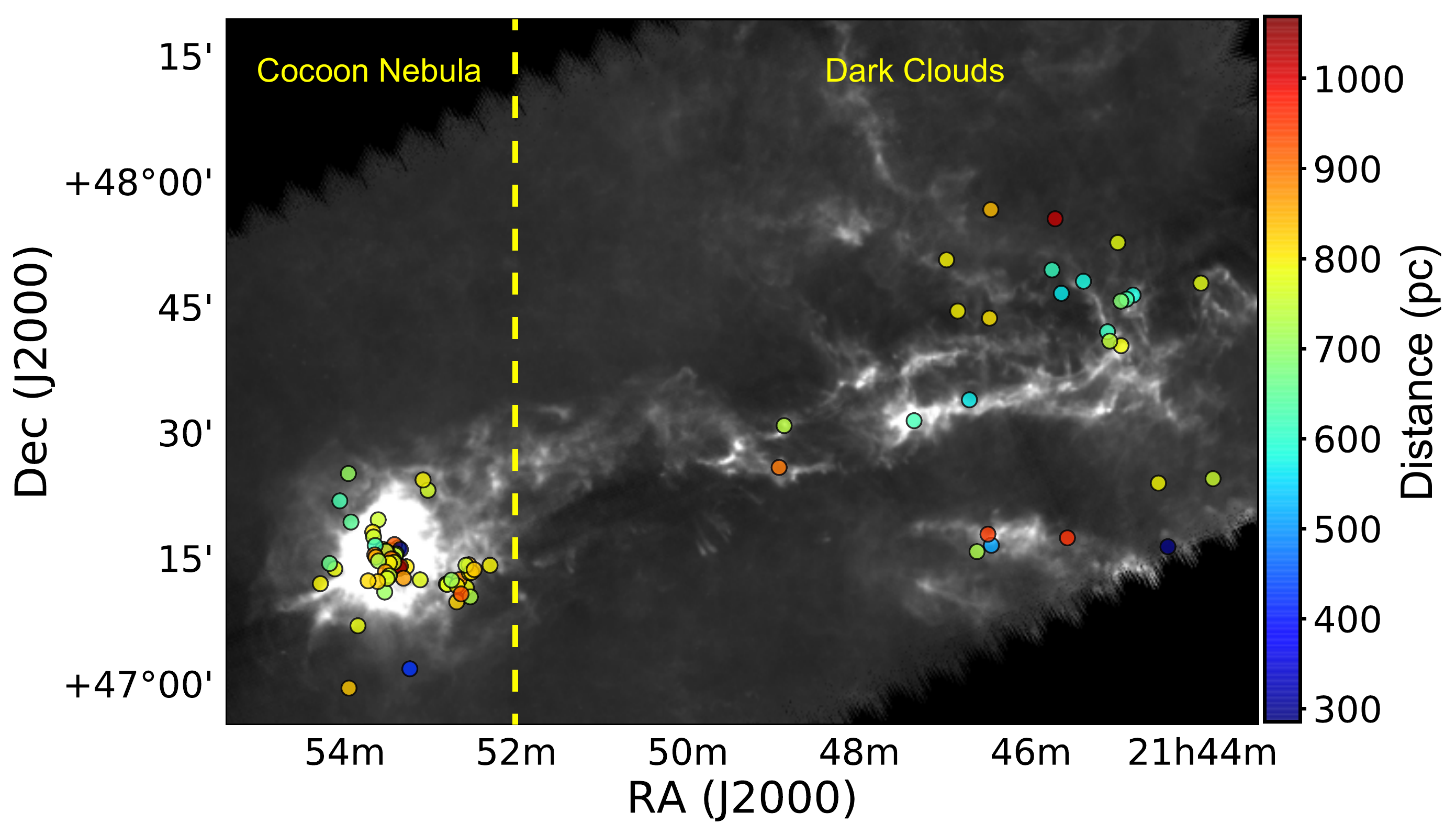}
\caption{The YSOs with distance measurements in the IC5146 cloud system. The YSOs were identified in \citet{ha08} and the color of each filled circle represents its parallax distance estimated from the $Gaia$ DR2 catalog. The yellow dashed line marks R.A. of $21^h52^m$ which we used to separate the YSOs. Most of the YSOs within the Cocoon Nebula have distances of $\sim 800$ pc, but the distances of the YSOs in the dark clouds are widely distributed from 300 to 1100 pc. The massive western main filament likely has a distance of $\sim 600$ pc, and is thus probably not physically associated with the Cocoon Nebula.}\label{fig:distance_map}
\end{figure*}

We used $Gaia$ DR2 data to perform a distance estimation toward both the Cocoon Nebula and the dark clouds. We selected YSOs, identified by \citet{ha08}, within the IC5146 sky area, and found the corresponding parallax measurements in the $Gaia$ catalog. Only those YSOs with significant parallax measurements ($>5\sigma$) were used. We further excluded 5 YSOs showing very large distances (1.1--2.6 kpc), out of a total of 98 YSOs, because they are possibly misidentified background sources. The surviving YSOs were separated into two groups, based on their spatial distributions: those in the Cocoon Nebula (R.A. $> 21^h52^m$) and those in the dark clouds (R.A. $< 21^h52^m$).

\begin{figure}[hbtp]
\includegraphics[width=\columnwidth]{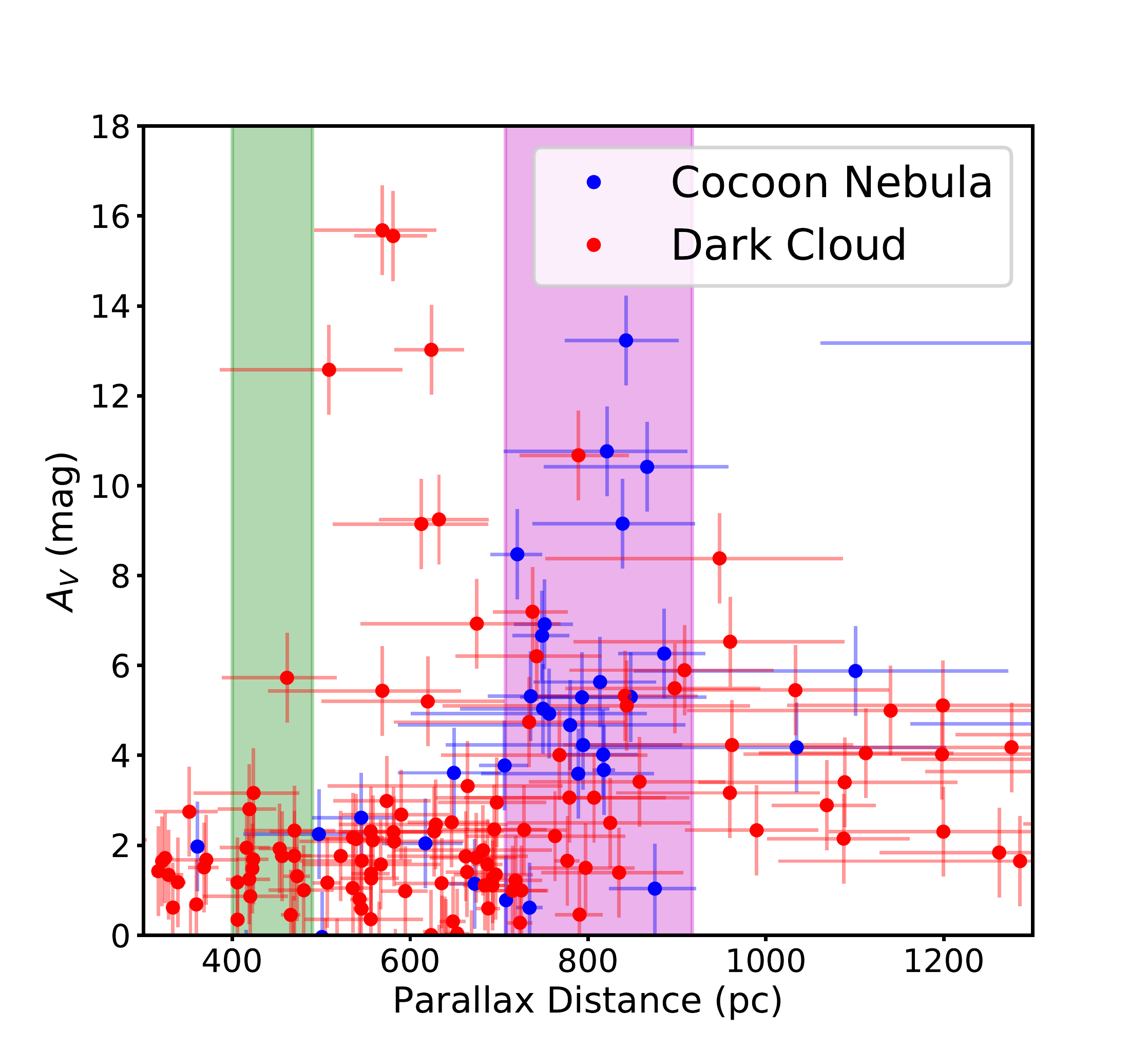}
\caption{Visual extinction versus parallax distance of stars within the IC5146 dense regions. In the Cocoon Nebula, the $A_V$ of stars rises at a distance of $\sim$700 pc, consistent with the distance estimation of $813\pm106$. However, many stars in the dark clouds with $A_V > 3$ mag are found at distances of 500--700 pc, suggesting that at least part of the dark cloud is at a close distance. The distance estimation of $460\substack{+40 \\ -60}$ pc is not supported by the $Gaia$ data.}\label{fig:D_AV}
\end{figure}

\autoref{fig:histo_d} shows the parallax distance measurements toward the YSOs in the Cocoon Nebula and in the dark clouds. The YSOs in the Cocoon Nebula have distances ranging from 300 to 1100 pc and show one major peak at a distance of $\sim$800 pc, consistent with previous results \citep{ha08,dz18}. In contrast to that, the YSOs in the dark cloud show two components, one with a peak at a distance of $\sim$800 pc similar to the Cocoon Nebula, and the other with a peak at a distance of $\sim$600 pc.

In order to illustrate the distance distributions of these YSOs, we label the YSOs with their estimated distance overlaid on the $Herschel$ 250~$\mu$m map in \autoref{fig:distance_map}. Most of the stars near the center of the Cocoon Nebula have similar distances of $\sim800$ pc. However, the YSOs around the dark clouds have a wide range of distances; the YSOs within the densest filament have similar distances of $\sim600$ pc, and the YSOs in the more diffuse areas seem to have distances of $\sim800$ pc. Because of the relatively small numbers of YSOs distributed over the dark clouds, parts of the dark clouds still lack distance information.

In order to examine the distance distribution of the clouds with more samples, we selected stars, instead of only YSOs, from the $Gaia$ DR2 catalog within the regions where the column density, estimated using the \textit{Herschel} five-band data, was greater than $3\times 10^{21}$ cm$^{-2}$ ($\sim A_V > 3$ mag). These stars were matched to extinction estimates from Paper \Rom{1}. \autoref{fig:D_AV} shows $A_V$ versus parallax distance for these 222 stars and YSOs. Since the uncertainty of the Paper \Rom{1} extinction estimates is $\sim 1$ mag, the stars with $A_V < 2$ mag could possibly be foreground stars. In the Cocoon Nebula, the $A_V > 2$ mag stars are consistent with distances of $\sim800$ pc. In the dark clouds, several $A_V > 10$ mag stars appear at distances of 500-700 pc. These correspond to the location of the densest filament in the dark cloud. These stars suggest a distance of $600\pm100$ pc, for the densest filament. We note that this analysis does not rule out the possibility that parts of the dark clouds may be more distant. 

Our results suggest that what is called the IC5146 cloud likely consists of two separate clouds. The first cloud is at a distance of $600\pm100$ pc, and contains at least the densest $A_V > 10$ mag main filaments. The second cloud is at a distance of $800\pm100$ pc, and is associated with the Cocoon Nebula. Nevertheless, we do not have sufficient information to show to which cloud the remainder of the filaments belong. In this paper, we adopt a distance of $600\pm100$ pc for the dark cloud and a distance of $800\pm100$ pc for the Cocoon Nebula for consistency.

\subsection{Magnetic Field Morphology}
\label{sec:B_morphology}

\begin{figure*}[hbtp]
\includegraphics[width=\textwidth]{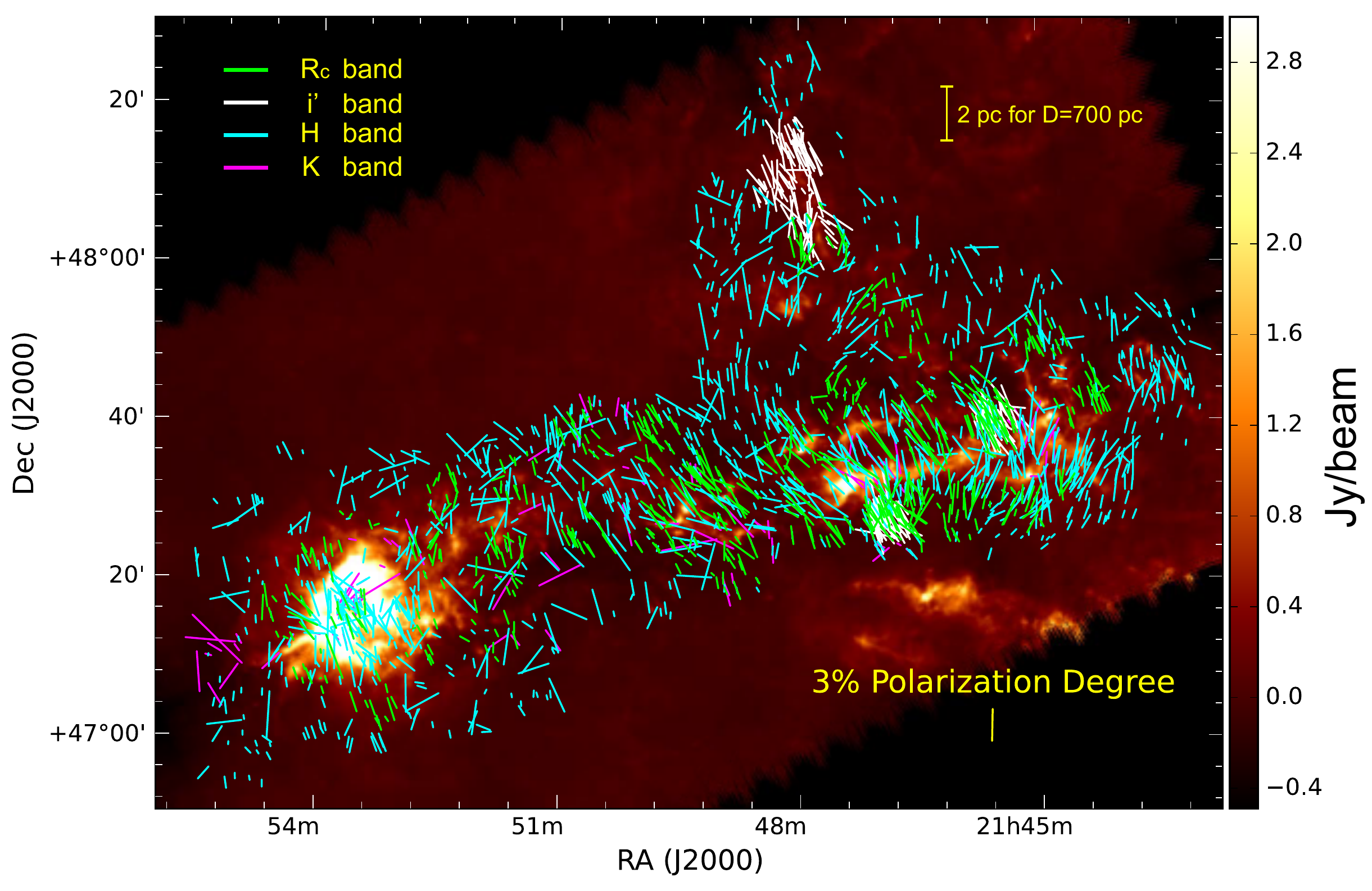}
\caption{Map of IC5146 stellar polarizations overlaid on a false-color representation of the \textit{Herschel} 250~$\mu$m intensity image. The polarization detections in TRIPOL $\mathrm{i'}$-band, AIMPOL $\mathrm{R_c}$-band, Mimir H- and K-band, taken from from Paper \Rom{1}, are shown as pseudovectors with white, green, cyan, and magenta colors.}\label{fig:pmap}
\end{figure*}

\begin{figure*}[hbtp]
\includegraphics[width=\textwidth]{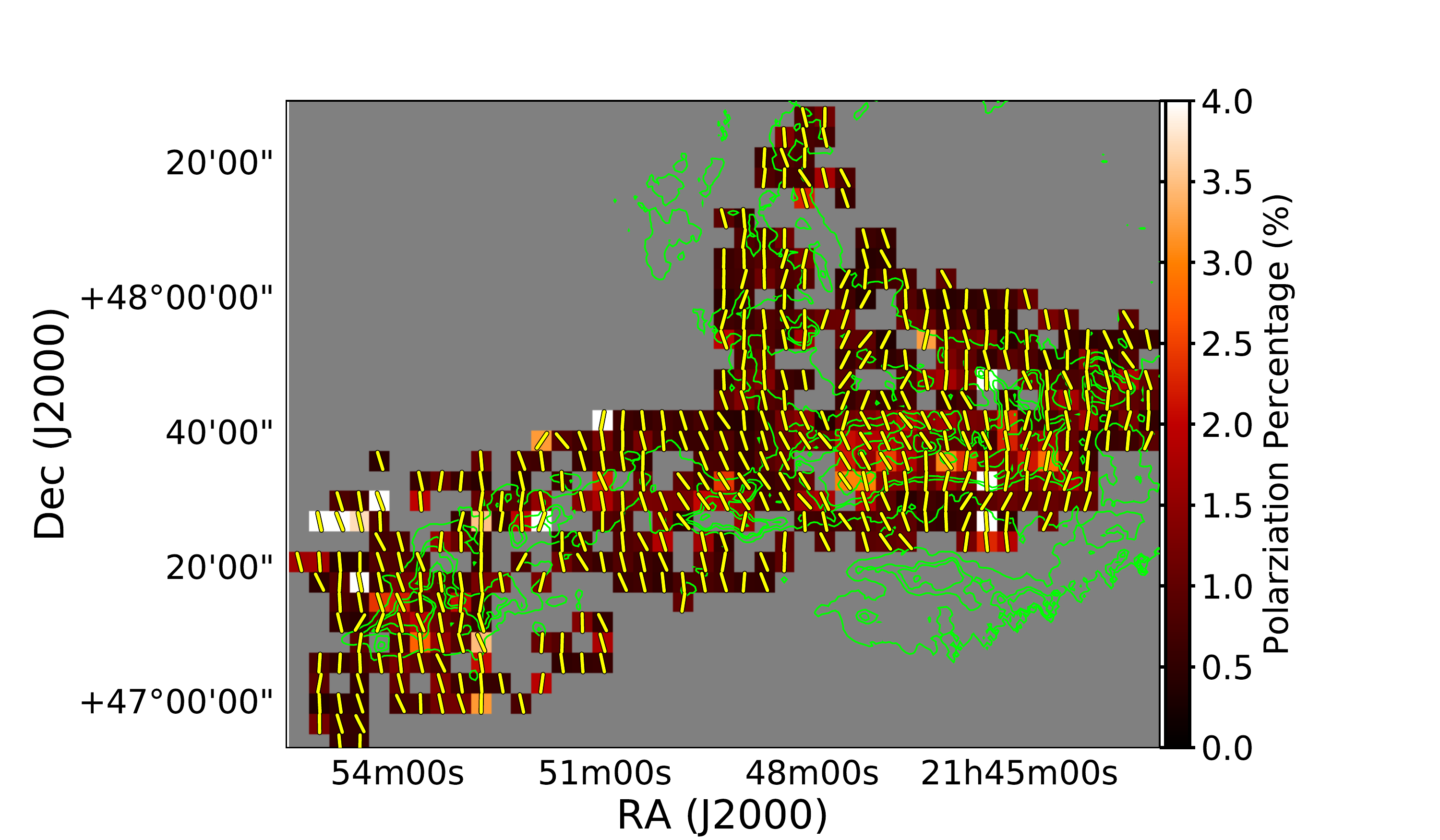}
\caption{Map of spatially-averaged H-band stellar polarization detections over the IC5146 cloud, using $3\times3$ arcmin pixels. The false-color representation of each pixel shows the polarization percentage. The yellow segments represent the orientation of the polarization. Green contours are H$_2$ column density, with levels of 1, 2, 3, and 10$\times$(10$^{21}$) cm$^{-2}$ calculated by \citet{ar11} using \textit{Herschel} data. Average polarization pixels are only shown if their S/N is greater than three.}\label{fig:s_pmap}
\end{figure*}

In Paper \Rom{1}, we reported optical and near-infrared starlight polarization over the IC5146 filamentary clouds shown in \autoref{fig:pmap}. The spatial distribution of starlight polarization detections is not uniform and highly depends on the cloud extinction, background star density, and observational conditions. The uneven sampling may bias quantities derived from the polarization patterns due to heavier weighting for regions where more detections are present. In order to minimize the uneven sampling effects, we generated a spatially-averaged H-band polarization map, with $3\times3$ arcmin bins. This reveals the magnetic field structure on a 0.6 parsec-scale. We note that this averaging process only helps to reduce the uneven sampling effects for scales greater than the pixel size, and the fundamental lack of polarization information in the dense regions can only be improved by additional data.

The averaged polarization map was calculated using all of the H-band polarization data in Paper \Rom{1}, including data with low S/N. In order to remove possible foreground stars, we rejected stars with polarization degree less than 0.3\%, which is the upper limit of foreground polarization. We calculated the inverse-variance weighted mean Stokes Q and U of the background stars over each $3\times3$ arcmin grid zone, and then computed the mean debiased polarization degree and position angle (PA). Only pixels where the polarization values were greater than three times their propagated uncertainties are shown in the map. 

The spatially-averaged polarization map is shown in \autoref{fig:s_pmap}. The map indicates an overall, several parsec-scale organized magnetic field that appears to have small pixel-to-pixel angular dispersion. In general, the magnetic field is perpendicular to the main filament, and parallel to the sub-filament that extends to the north. The magnetic field in the western part of the cloud likely shows a large-scale curvature, where the polarization PA changes from $-20$\degr\ to 20\degr\ over $\sim$4 pc. 

\begin{figure*}[hbtp]
\includegraphics[width=\textwidth]{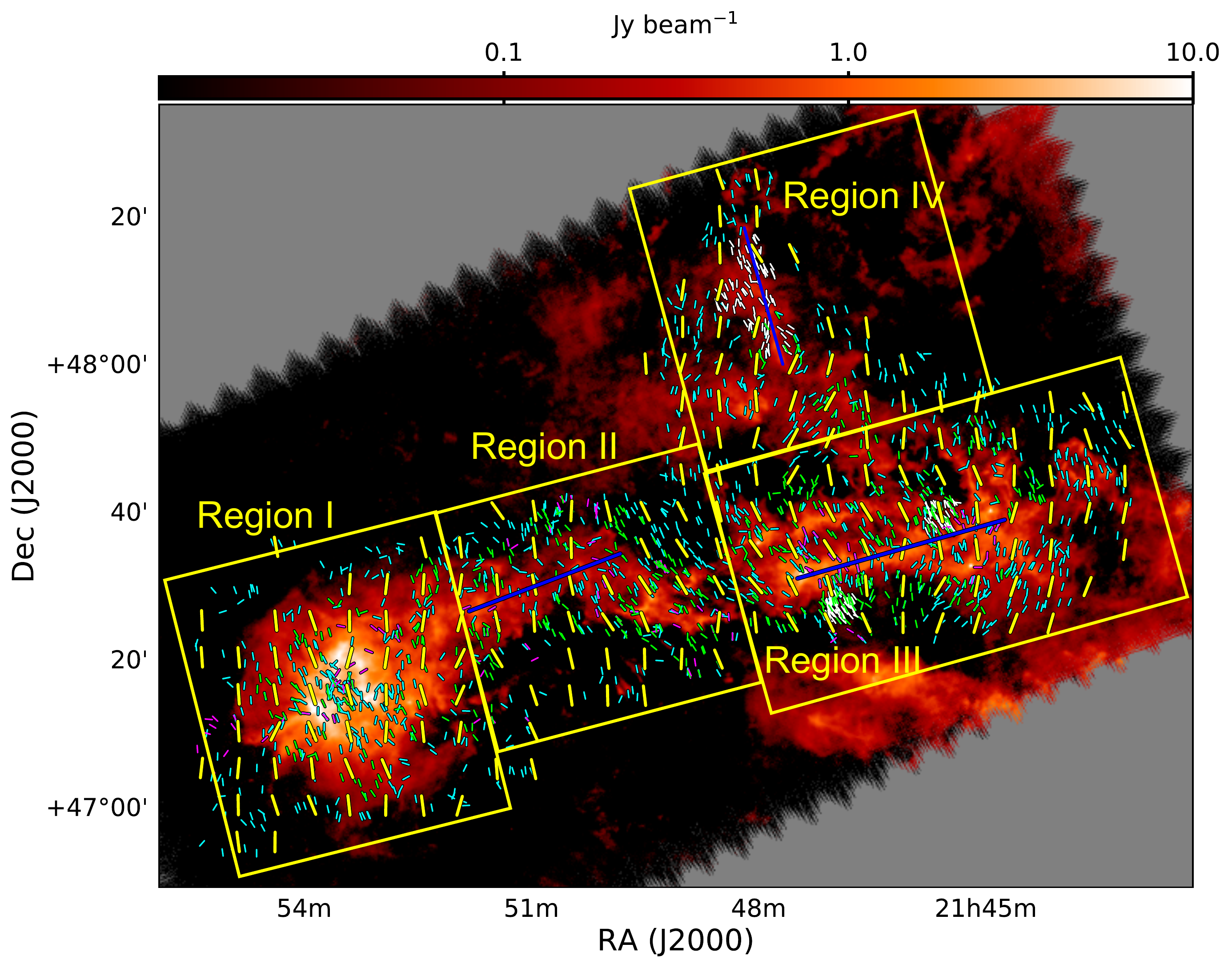}
\caption{Selected region zones, with the stellar polarization map overlaid on the $Herschel$ 250~$\mu$m map. The detections in $\mathrm{i'}$-, $\mathrm{R_c}$-, H-, and K-bands are shown with white, green, cyan, and magenta segments, respectively. The thicker yellow segments are the spatially-averaged H-band polarization detections. The yellow boxes identify the four sub-regions with different evolutionary stages. The dark blue lines are the filament ridges which we used to calculate the projected distance in \autoref{fig:DPA}}\label{fig:region_pmap}
\end{figure*}

The IC5146 cloud consists of filaments in a variety of evolutionary stages, identified by the hosting YSO populations and filament stabilities \citep{ha08,ar11,jo17}. We separated the system into four sub-regions, as shown in \autoref{fig:region_pmap}. The regions include (1) the Cocoon Nebula H\Rom{2} region, (2) the Eastern Main Filament, (3) the Western Main Filament, and (4) the Northern Filament that extends from the main filament structure. The Cocoon Nebula and the western part of the main filament are both star-forming regions \citep{ha08}, and the Cocoon Nebula is likely more evolved because it displays a lower fraction of Class 0/\Rom{1} YSOs in the full population \citep{ha08,jo17}. On the other hand, the eastern main filament and the northern filament are quiescent and may be gravitationally stable, as no YSOs are present \citep{ar11}.

\begin{figure*}
\includegraphics[width=\textwidth]{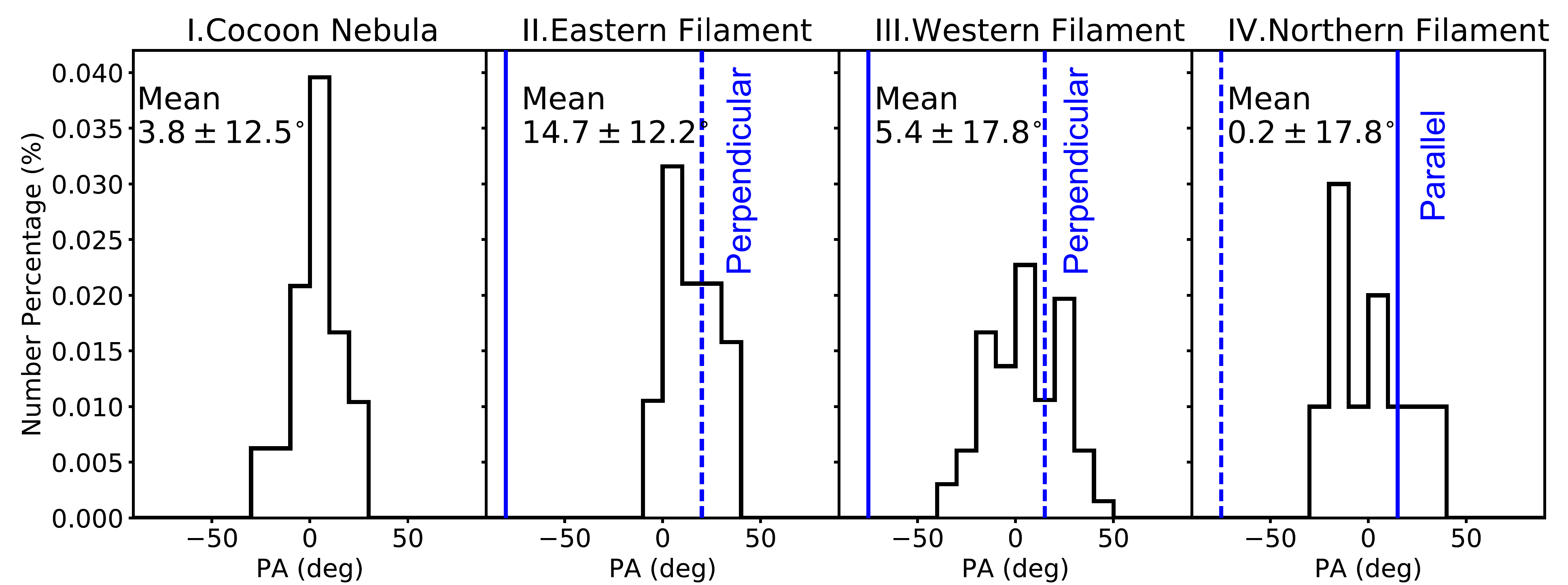}
\caption{Histograms of smoothed H-band polarization PAs toward the four regions delineated in \autoref{fig:region_pmap}. The blue solid and dashed lines show the PAs that are parallel and perpendicular to the main filament, respectively. The mean magnetic field orientations in these four regions are similar to within $\sim10\degr$. The angular dispersion of $\sim12\degr$ in the eastern side is less than the angular dispersion of $\sim18\degr$ in the western side. Inset mean values are indicated with uncertainties representing the standard deviation of each distribution. }\label{fig:histo_PA}
\end{figure*}

To reveal the overall magnetic field morphologies of the four regions, PA histograms of the smoothed H-band grid data toward the four regions are shown in \autoref{fig:histo_PA}. The four regions have similar mean PA values, ranging from 0\fdg2 to 14\fdg8. The PA dispersions for the Cocoon Nebula and for the eastern main filament are both $\sim12\degr$, while the western main filament and the northern filament both have higher PA dispersions of $\sim18\degr$. Filament orientations, as plotted in \autoref{fig:region_pmap}, are marked in the histograms (\autoref{fig:histo_PA}) to show that the magnetic fields are mostly perpendicular to the eastern and western main filaments, but parallel to the northern filament.

\begin{figure*}
\includegraphics[width=\textwidth]{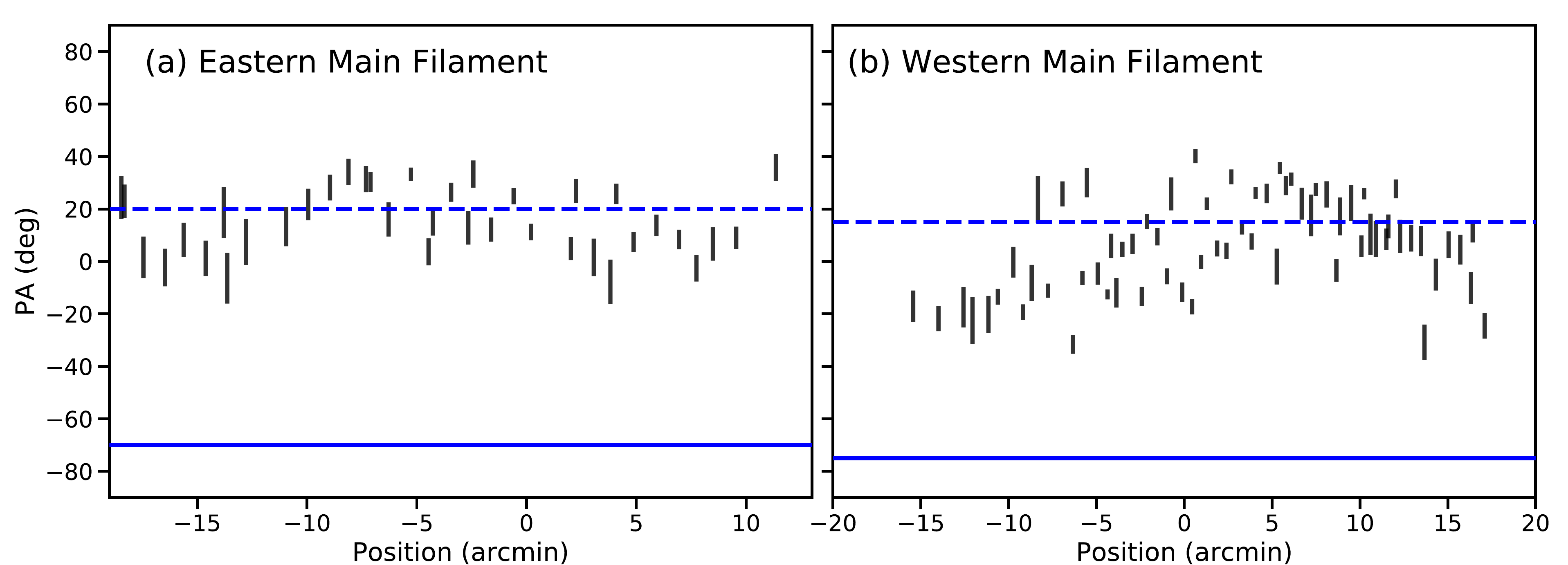}
\caption{Polarization PA versus projected distance to the filament ridge for (a) the Eastern Main Filament and (b) the Western Main Filament. The black lines show the spatially-averaged H-band polarization PA values and uncertainties. The positive offset direction is to northeast, perpendicular to the filament ridge. The blue solid and dashed lines show PA orientations parallel and perpendicular to the main filament, respectively. The magnetic field is mostly perpendicular to the eastern main filament, and the alignment does not change with projected distance. (b) The magnetic field orientation systematically varies from $\sim-20\degr$ to $\sim0\degr$ as the projected distance changes from 5--20 to $<0$ and $>20$ arcmin, where the magnetic field becomes slightly misaligned to the main filament. This variation indicates a large-scale curved magnetic field morphology, possibly caused by the contraction along the massive filament.} \label{fig:DPA}
\end{figure*}

In order to investigate whether the higher PA dispersion in the western main filament is caused by large-scale structured changes, we plot PA versus the projected distance to the filament ridge in \autoref{fig:DPA}. The filament ridge (zero point of the projected distance) is identified by eye, as shown in \autoref{fig:region_pmap} as dark blue lines, and the positive offset direction is to the northeast, perpendicular to the filament ridge. In the eastern main filament, the PA shows no correlation with the projected closest distance and is always perpendicular to the filament orientation (20\degr), to within 10\degr. In contrast, for the western filament the PA systematically increases from $-20$\degr\ at a projected distance of $-15$~arcmin to 20\degr\ at a projected distance of $-5$~arcmin, and then back to $-20$\degr\ when the projected distance is greater than 10~arcmin. The magnetic field is perpendicular to the main filament, to within 10\degr, when the projected distance is between $-5$ to 10~arcmin, but the PA shows a $\sim30\degr$ offset to the filament orientation when the projected distance is outside that range. This feature is consistent with the large-scale curvature noted previously in the polarization map, and likely causes the higher PA dispersion measured for this filament.

\subsection{Magnetic Field Strength over the IC5146 Cloud}\label{sec:B_str}
The Davis-Chandrasekhar-Fermi (DCF) method \citep{da51,ca53} is commonly used to estimate the strength of the plane-of-sky component of the magnetic field ($B_{pos}$) using dust polarization data \citep[e.g.,][]{hi09,ma12,ho17}. Assuming that the turbulent kinetic energy and the magnetic energy are in equipartition, the DCF method suggests that $B_{pos}$ is estimated using
\begin{equation}\label{eq:CF}
B_{pos}=Q~\sqrt[]{(4\pi \rho)} \frac{\sigma_v}{\delta\phi}, 
\end{equation} 
where ${\delta}{\phi}$ is the dispersion of measured polarization orientations, $\sigma_v$ is the line-of-sight gas velocity dispersion, $\rho$ is the gas volume density, and $Q$ is a modification factor used to correct the overestimation of $B_{pos}$ due to possible complicated magnetic field structure along the line of sight. \citet{os01} found $Q=0.5$ yielded a good approximation of their MHD simulations if the magnetic field angular dispersions were less than 25\degr.

\subsubsection{Polarization Angle Dispersions}\label{sec:disp}
To apply the DCF method, we estimated the polarization position angular dispersions across the IC5146 cloud. The DCF method assumes that the magnetic field angular dispersion is only due to the response of the field to gas turbulence; however, in reality, magnetic fields often have a non-uniform geometry over large-scales. To remove large-scale PA patterns, we used a $6\times6$~arcmin grid of bins in which we calculated local mean polarization PAs and their corresponding angular dispersions. Each bin serves as a boxcar filter that blocks the PA patterns on scales greater than the box size (1.4 pc), and the bin size was chosen to sufficiently filter out the $\sim4$ parsec-scale curvature shown in \autoref{sec:B_morphology}. The grid-based calculation ensures that each stellar polarization detection was used in only one bin, so that the estimates among bins are independent.

\begin{figure}
\includegraphics[width=\columnwidth]{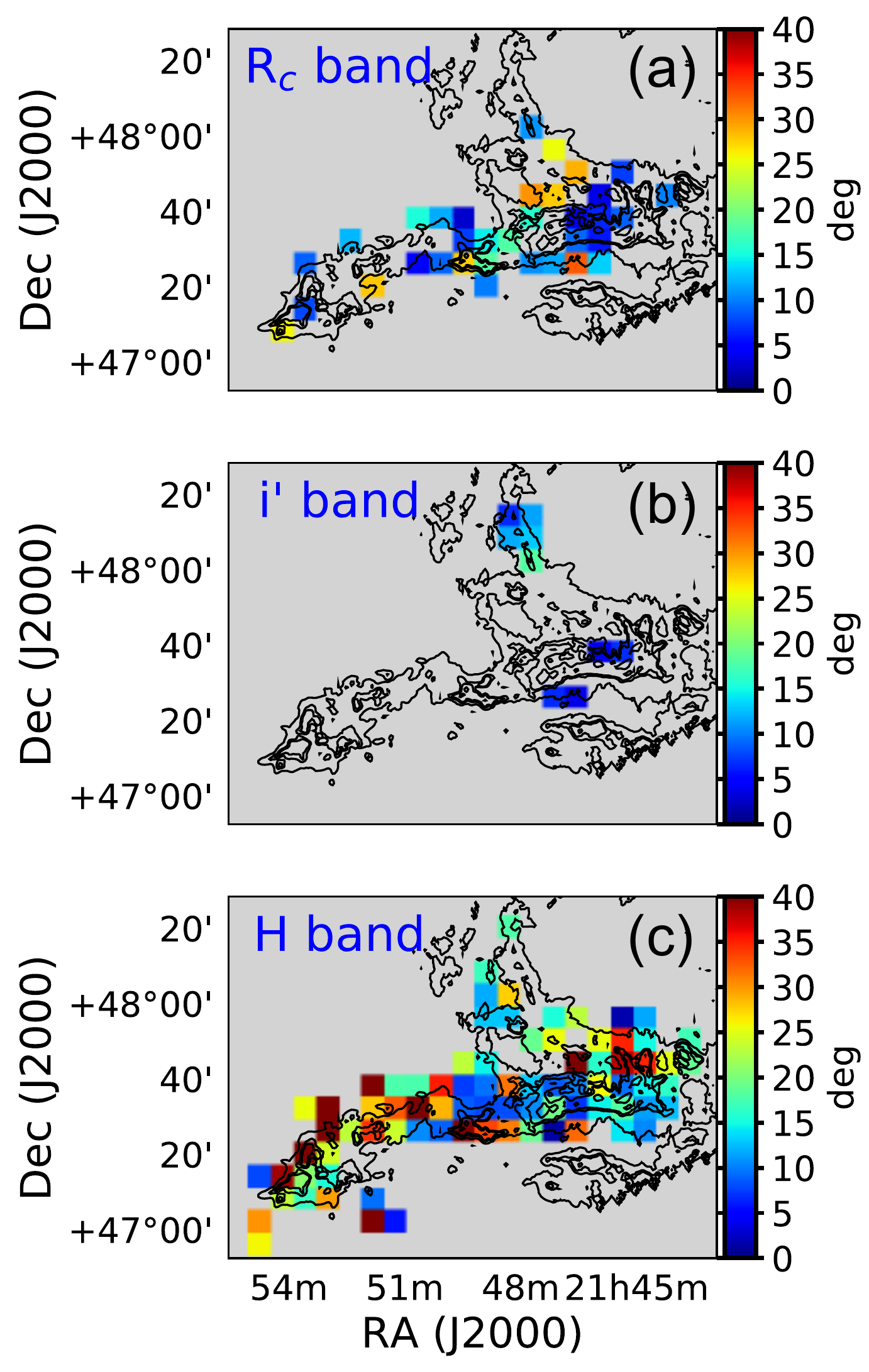}
\caption{PA dispersion maps for $\mathrm{R_c}$-, $\mathrm{i'}$-, and H-band data, calculated using $6\times6$ arcmin bins. The black contours are the \textit{Herschel} 250~$\mu$m intensities, at levels of 0.1, 1, and 5 Jy/beam.}\label{fig:s_dismap}
\end{figure}

Only the stars with S/N$ > 3$ for $\mathrm{R_c}$- and $\mathrm{i'}$-band or S/$N>2$ for H-band were used to calculate the angular dispersions. The angular dispersions were calculated using the $\mathrm{R_c}$-, $\mathrm{i'}$-, and H-band data separately, because they may trace different parts of the cloud. The angular dispersions calculated using less than 7 PA values were excluded. The inverse-variance weighted standard deviation of the PA distribution was calculated in each bin and corrected for instrumental uncertainty by 
\begin{equation}
\delta\phi = \sqrt{\frac{\sum\limits_{i=1}^{n}((\mathrm{PA}_i - \overline{\mathrm{PA}})^2-\sigma_{\mathrm{PA_i}}^2)/\sigma_{\mathrm{PA_i}}^2}{\frac{n-1}{n}\sum\limits_{i=1}^{n}1/\sigma_{\mathrm{PA_i}}^2}},
\end{equation}
where $\mathrm{PA}_i$ and $\sigma_{\mathrm{PA_i}}$ are the observed PA and its uncertainty for the $i$th segment, and $\overline{\mathrm{PA}}$ is the inverse-variance weighted mean $\mathrm{PA}$ in each selected bin. To handle the $\pm$180 degree ambiguity, the angular dispersions were calculated in different PA coordinate systems, and only the minimum dispersions were used. The resulting angular dispersion maps are shown in \autoref{fig:s_dismap}. The mean number of PAs used to calculate the angular dispersions are 10, 18, and 13 for the $\mathrm{R_c}$-, $\mathrm{i'}$-, and H-bands, respectively. The mean propagated uncertainties in the angular dispersions are 1\fdg1, 0\fdg4, and 3\fdg4 for the $\mathrm{R_c}$-, $\mathrm{i'}$-, and H-bands. The mean S/N of the angular dispersions are 13.5, 25.1, and 8.4 for the $\mathrm{R_c}$-, $\mathrm{i'}$-, and H-bands. 

\subsubsection{H$_2$ Volume Density}\label{sec:volden}
The IC5146 cloud consists of numerous filamentary clouds. \citet{ar11} identified 27 filaments in this system based on the \textit{Herschel} data, and showed that the density structure of these filaments could be well described by the Plummer-like profile:
\begin{equation}\label{eq:volden}
n_p(r)=\frac{n_c}{[1+(r/R_{flat})^2]^{p/2}}
\end{equation}
where $n_c$ and $n_p(r)$ are the $H_2$ volume density at the filament ridge and at a radial offset distance $r$ to the filament ridge. The profile parameter $R_{flat}$ represents the scaling radius of the column density profile, and $p$ is the profile index representing the mass concentration of filaments. The observed values of $R_{flat}$ and $p$ are in range of $\sim0.02$--0.06 pc and $\sim1.5$--2.1 in the IC5146 cloud \citep{ar11}.

\begin{figure}
\includegraphics[width=\columnwidth]{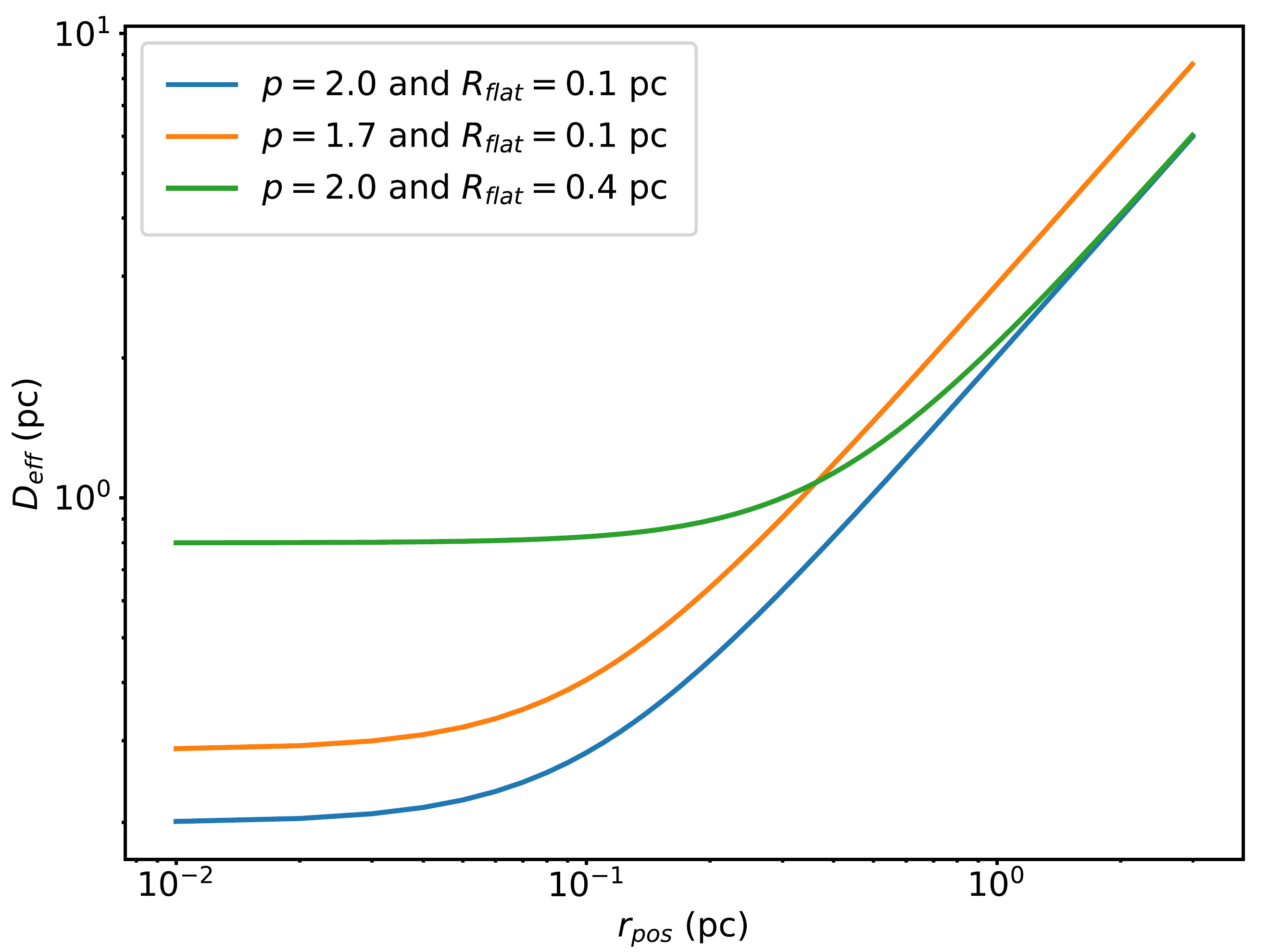}
\caption{The effective thickness $D_{eff}$ of Plummer-like profiles defined in \autoref{eq:thick}. The x-axis $r_{pos}$ represents the projected radial offset in the plane of the sky. The effective thickness is approximately proportional to $r_{pos}$ if $r_{pos} >> R_{flat}$.}\label{fig:thick}
\end{figure}

To estimate the mean volume density along the line of sight, a cloud boundary is required. However, the Plummer-like profile describes a structure extending to infinite $r$, and thus provides no well-defined boundary. Here we defined an effective thickness ($D$) of a Plummer-like structure as the thickness of the central regions that contributes half of the total column density, as
\begin{equation}\label{eq:thick}
\frac{\int_{\frac{-D}{2}}^{\frac{D}{2}}n_p(r)dr_{los}}{\int_{-\infty}^{\infty}n_p(r)dr_{los}}=\frac{1}{2},
\end{equation}
where $r_{los}$ is the line of sight component of the radial offset.
If the Plummer profile parameters $R_{flat}$, $p$, and the projected radial offset $r_{pos}$ are given, the effective thickness could be computed numerically using \autoref{eq:volden} and \autoref{eq:thick}.
\autoref{fig:thick} shows the computed effective thickness ($D$) versus $r_{pos}$ for various $p$ and $R_{flat}$. The effective thickness is approximately proportional to $r_{pos}$ when $r_{pos} >> R_{flat}$. The mean volume density along the line of sight could be estimated by
\begin{equation}\label{eq:mean_den}
\overline{n}= \frac{N_{obs}}{2D},
\end{equation}
where $N_{obs}$ is the observed total column density.

\begin{figure}
\includegraphics[width=\columnwidth]{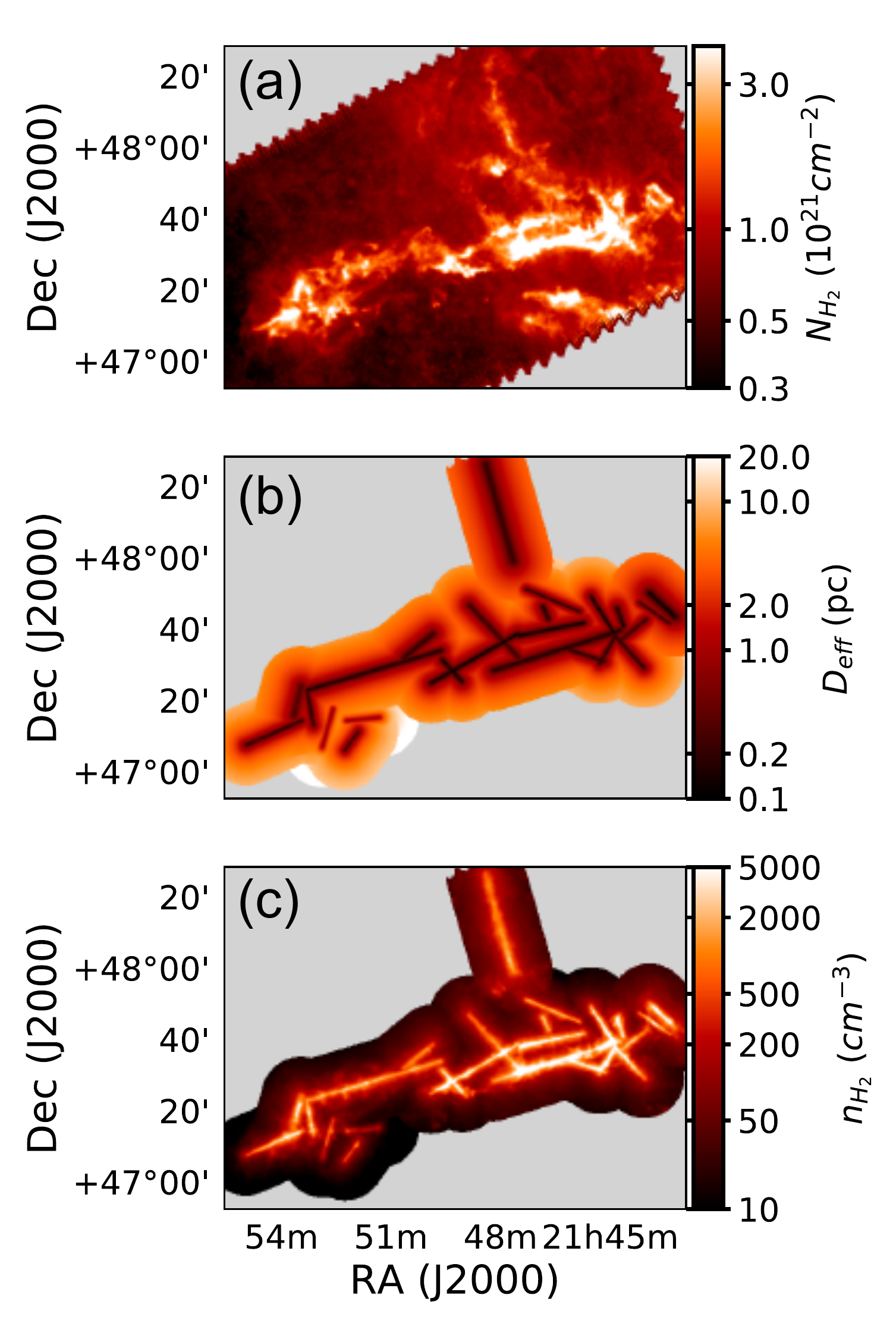}
\caption{(a) Column density map for the IC5146 cloud, with a beam size of 35~arcsec, calculated by \citet{ar11} using \textit{Herschel} data. (b) Effective thickness map computed using \autoref{eq:thick} with the Plummer parameters of the 27 filaments identified by \citet{ar11}. (c) Volume density map computed by the column density map and the Plummer effective thickness method (\autoref{eq:volden}). }\label{fig:volden}
\end{figure}

In order to create a volume density map using \autoref{eq:mean_den}, a column density map and an effective thickness map were required. Here we assumed that the filaments in the IC5146 cloud were cylindrical and following the Plummer-like profile to estimate the mean volume densities. We used the $H_2$ column density map of \citet[][see \autoref{fig:volden}a]{ar11} with a beam size of 35~arcsec, calculated from the \textit{Herschel} five bands data. We adopted the locations and the fitted Plummer parameters of the 27 filaments identified by \citet{ar11} and labeled these filaments on the column density map. For each pixel in the volume density map, we calculated the projected radial offsets to each of the 27 filament ridges as $r_{pos}$. If the calculated $r_{pos}$ was less than 11~arcmin, the boundary that \citet{ar11} used to fit the filament density profile, we further computed the effective thickness using \autoref{eq:thick} with the $r_{pos}$, and the Plummer parameters $R_{flat}$ and $p$ associated with the filament. The computed effective thickness was assigned to the pixel to create an effective thickness map. 

For areas covered by more than one filament, we assumed that the filament network was fragmented from the same cloud, and so these neighboring filaments were mostly spatially-overlapped. Hence, if more than one effective thickness value was assigned to a pixel, the lowest value was used, since filaments with higher local density are expected to contribute more polarized intensity. Because \citet{ar11} derived the Plummer parameters assuming a cloud distance of 460 pc, we scaled the derived thicknesses to values appropriate for distances of $600\pm100$~pc and $800\pm100$~pc for the filaments located in the dark clouds (R.A.$< 21^h52^m$) and in the Cocoon Nebula (R.A.$> 21^h52^m$), respectively. The resulting effective thickness map is shown in \autoref{fig:volden}(b), and is used to create the mean volume density map, as shown as \autoref{fig:volden}(c), using \autoref{eq:mean_den}.

To estimate the mean volume densities that match our position angle dispersion map on the same $6\times6$~arcmin grid, the mean volume density map needed to be smoothed. Because our polarization data cannot trace the densest regions, directly averaging of the densities of all pixels may include regions in which we have no polarization data. To avoid this bias, we only selected the volume density values for the locations of the polarization detections that we used to calculate the position angle dispersions. The selected density values, for the $\mathrm{R_c}$-, $\mathrm{i'}$-, and H-band polarization detections, were averaged within the $6\times6$~arcmin grid separately, and created three mean volume density maps, each one corresponding to the $\mathrm{R_c}$-, $\mathrm{i'}$-, and H-band data.

\subsubsection{Gas Velocity dispersion}\label{sec:vdisp}
\citet{ar13} measured $^{13}$CO (2-1), C$^{18}$O (2-1), and N$_{2}$H$^{+}$ (1-0) lines toward several filaments in the IC5146 cloud. Although their observations only consisted of a single pointing toward each of the filaments, the observed lines show roughly constant non-thermal velocity dispersions of $0.20$ km~s$^{-1}$ with a standard deviation of 0.06 km~s$^{-1}$, if the column density is $\lesssim10^{22}$ cm$^{-2}$. This standard deviation is much greater than their observational uncertainties of 0.01 km~s$^{-1}$, and thus likely represents the intrinsic difference among these filaments. Because most of our polarization detections are located in $A_V\lesssim10$ mag regions, we assume a constant velocity dispersion of 0.20$\pm$0.06 km~s$^{-1}$.

\subsubsection{Magnetic Field Strengths}\label{sec:Bmap}
\begin{figure}
\includegraphics[width=\columnwidth]{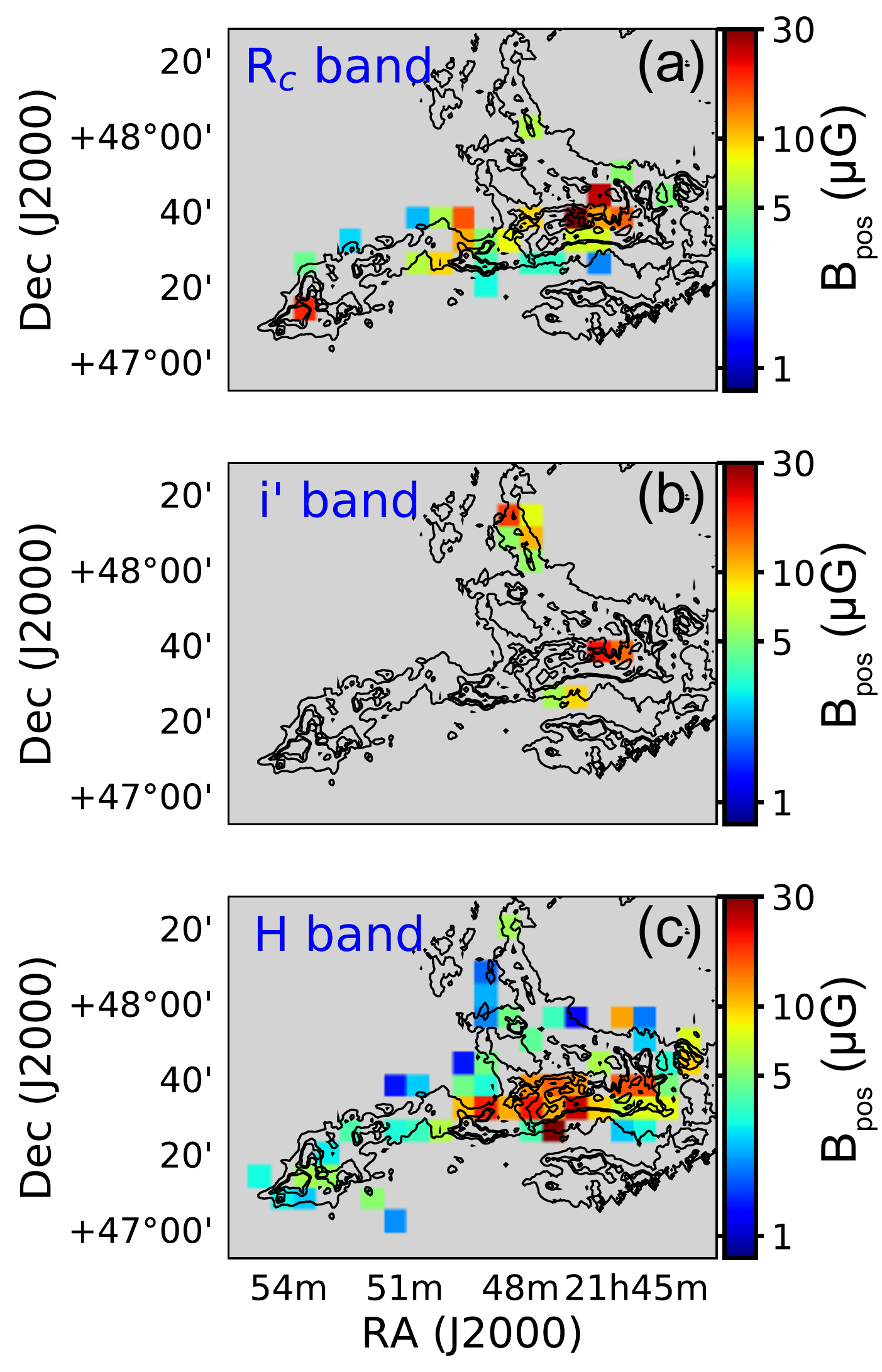}
\caption{Plane-of-sky magnetic field strength, $B_{pos}$, maps estimated using the $\mathrm{R_c}$-, $\mathrm{i'}$-, and H-band binned data. Only 6$\times$6 pixels with PA dispersion ($\delta \phi) <$ 25\degr\ were used to estimate the magnetic field strength. The black contours are the \textit{Herschel} 250~$\mu$m intensities, at levels of 0.1, 1, and 5 Jy/beam. }\label{fig:Bmap}
\end{figure}

With the above estimated quantities, we calculated $B_{pos}$ using \autoref{eq:CF} for each of the $\mathrm{R_c}$-, $\mathrm{i'}$-, and H-band data sets separately. For those pixels with $\delta \phi > 25\degr$, the magnetic field strength could not be accurately estimated using \autoref{eq:CF}, as per \citet{os01}. About 30\% of the pixels have $\delta \phi > 25\degr$, and these pixels were excluded. The resulting magnetic field strength maps are shown in \autoref{fig:Bmap}. 

The $\mathrm{R_c}$ and H-band data mostly trace the main filament regions, while the $\mathrm{i'}$-band data mainly trace the northern regions. The estimated magnetic field strength ranges from a few $\mu$G in diffuse regions to 30 $\mu$G in the dense parts of the clouds. The median uncertainties of the magnetic field strength estimates are $33\%$, $31\%$, and $35\%$ for the $\mathrm{R_c}$-, $\mathrm{i'}$-, and H-bands, respectively. These uncertainties are dominated by the uncertainties in the velocity dispersion (30\%), which might be overestimated because the velocity dispersions were measured from only a small area of the filament ridges.

\subsection{Magnetic Field Strength versus Density}\label{sec:Bvsn}
\citet{cr10} analyzed the $B_{pos}$--$n$ relation revealed by the Zeeman effect measurements for numerous clouds, and found that (1) magnetic fields scale with volume density by a power-law index of 0.65, and (2) the PDF of the magnetic field strengths for clouds with similar densities is uniformly distributed. Both of these features favor the weak magnetic field star formation models. However, these findings might be biased by the diversity of samples taken from a variety of environments. In this section, we aim at testing \citet{cr10} using our samples that were obtained from similar environments.

\begin{figure}
\includegraphics[width=\columnwidth]{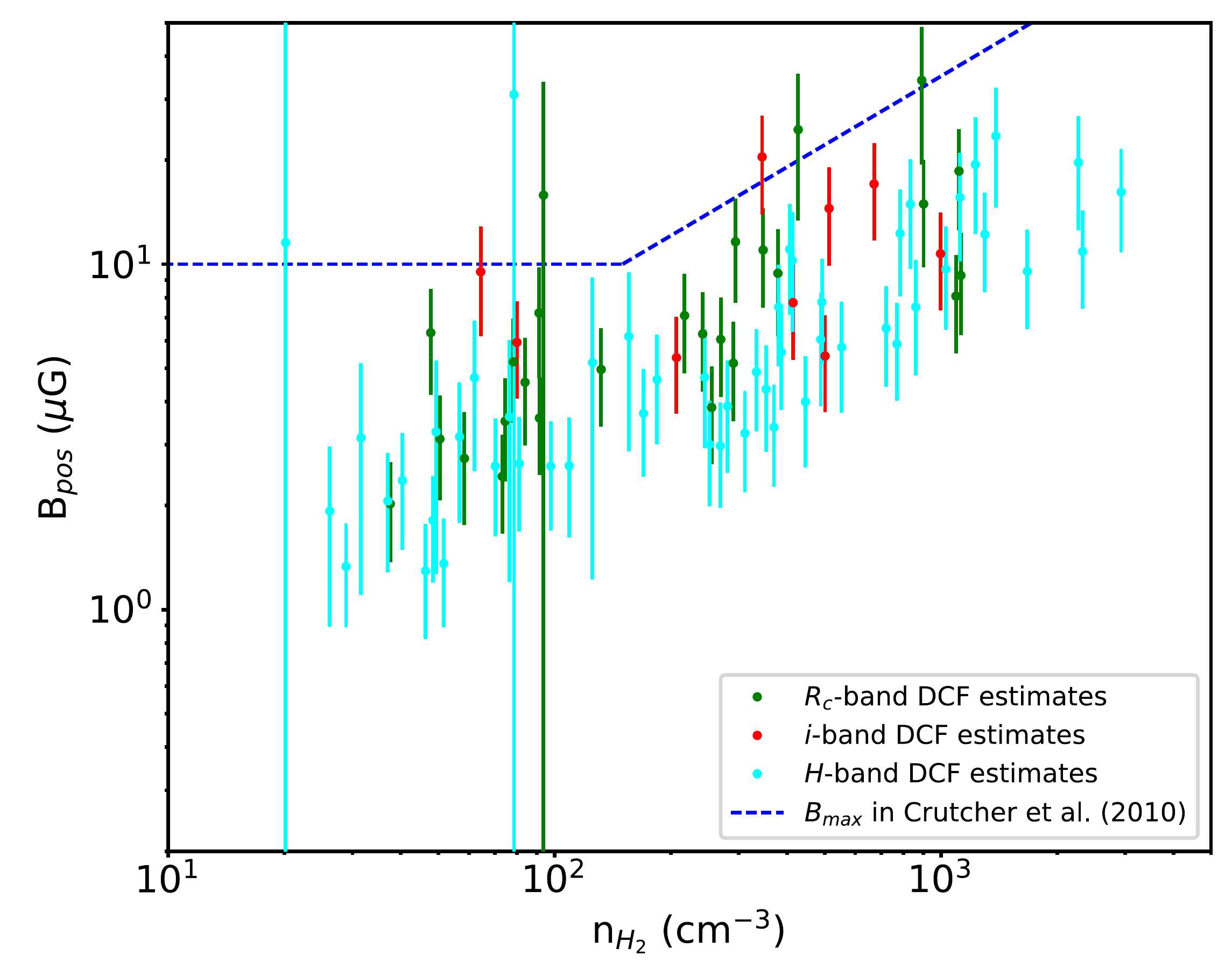}
\caption{$B_{pos}$ versus molecular hydrogen volume density $n$. The green, red, and magenta points represent the $B_{pos}$ and volume density values of the pixels for $\mathrm{R_c}$-, $\mathrm{i'}$-, and H-band binned data, respectively, as shown in \autoref{fig:Bmap}. Because all our samples were obtained in molecular gas, here we used $n_{H_2}$ volume density instead of $n_{H}$, which was used in \citet{cr10} for the H\Rom{1} cloud samples. The blue dashed line is the maximum magnetic field strengths obtained in \citet{cr10}, which shows a flat portion and another with a slope of 0.66. The $B_{pos}$--$n$ distribution of $\mathrm{R_c}$- and H-band samples are well-mixed, likely following similar trends. The $\mathrm{i'}$-band samples show a shallow slope, although the relation is uncertain due to the small number of samples.}\label{fig:Bvsn}
\end{figure}

We plot $B_{pos}$ versus cloud molecular hydrogen volume density $n$ in \autoref{fig:Bvsn} using the estimates based on $\mathrm{R_c}$-, $\mathrm{i'}$-, and H-band data. The $\mathrm{R_c}$- and H-band $B_{pos}$--$n$ distribution are likely well mixed and have similar trends. The $\mathrm{i'}$-band $B_{pos}$--$n$ distribution seems to have a less obvious slope because of the low number of samples. In addition, $\mathrm{i'}$-band data was mostly taken in the northern filament, where the magnetic field is parallel to the main filament, while the $\mathrm{R_c}$- and H-band data was mostly taken over the main filament, where the magnetic field is perpendicular to the main filament. Hence, the role of magnetic fields in regulating the cloud evolution traced by $\mathrm{i'}$-band data may be different from that traced by $\mathrm{R_c}$- and H-band data. As a result, here we only used the $\mathrm{R_c}$- and H-band data for analysis of the $B_{pos}$--$n$ distribution to reveal how magnetic fields regulate the evolution of the main filament.

To unveil the PDF of intrinsic magnetic field strength for cloud elements with the same volume density, and investigate how $B_{pos}$ scales with $n$, we used a Bayesian approach to analyze the observed $B_{pos}$--$n$ distribution. Bayesian statistics provides a framework for the quantitative comparison of models given data and an explicit set of assumptions. We tested the goodness of the models assuming either a uniform or a Gaussian PDF magnetic field strength and investigated the corresponding $B_{pos}$--$n$ power-law index. 

According to the Bayes' Theorem:
\begin{equation}
P(\theta|D)= \frac{P(D|\theta)P(\theta)}{P(D)}.
\end{equation}
The $P(\theta|D)$ (posterior probability) provides the PDF of the model, which is what we want to know. The $P(D|\theta)$ term is the likelihood function describing the probability of matching the observed data set $D$ using a given model parameter set $\theta$. The $P(\theta)$ term is our prior guess of the model parameters, and $P(D)$ is the distribution of data, which is a constant to normalize the probability function.

To determine the likelihood function $P(D|\theta)$, we assumed that the probability of observing the $i$th plane-of-sky magnetic field strength $B_{pos,i}$ is given by the convolution of the measurement probability function $P_{obs}(B_{pos,i},B)$ and the intrinsic magnetic field strength probability function $P_{int}({B,\theta}$) as 
\begin{equation}\label{eq:pdf}
P(B_{pos,i}|\theta)=\int P_{obs}(B_{pos,i},B)\times P_{int}({B,\theta})dB.
\end{equation}
Based on the magnetic field morphology shown in \autoref{sec:B_morphology}, we assumed that the magnetic field morphology in the IC5146 cloud is nearly uniform, and therefore $B_{pos}/B$ is a constant that does not change the PDF of $B_{pos}$. To represent our sample selection criterion $\delta \phi < 25\degr$, we added a boundary condition into the probability function by
\begin{equation}\label{eq:trunctation}
P(B_{pos,i}|\theta)=0 \qquad (B_{pos,i}< 0.5~\sqrt[]{(4\pi \rho)}. \frac{\sigma_v}{25\degr}).
\end{equation}
The likelihood function for the whole data set was the product of the probabilities of $N$ individual measurements:
\begin{equation}\label{eq:likelihood}
P(D|\theta)= \prod^{N}_{i=1} P(B_{pos,i}|\theta).
\end{equation}

Because the uncertainty of $B_{pos,i}$ are dominated by the uncertainty of the gas velocity dispersion, which is the standard deviation of the roughly constant velocity dispersions found in \citet{ar11}, here we approximated the $P_{obs}(B_{pos,i},B)$ by a Gaussian function as
\begin{equation}
P_{obs}(B_{pos,i}|B)=G(B_{pos,i},B,\sigma_{B_{pos,i}}),
\end{equation}
where $G(x,\mu,\sigma)$ denotes the Gaussian function, and $\sigma_{B_{pos,i}}$ is the propagated uncertainties of $B_{pos,i}$.

The intrinsic magnetic field strength probability function $P_{int}({B,\theta})$ is determined by our $B$--$n$ model. Here we tested two different model sets: (1) cloud elements with similar densities have uniformly distributed magnetic field strengths, representing a cloud with random and scattered mass-to-flux ratios and favoring the weak field star-formation theories, or (2) cloud elements with similar densities have Gaussian distributed magnetic field strengths, indicating a cloud with narrowly distributed mass-to-flux ratios. The setup of these two models is described in the following sections.

\subsubsection{Gaussian PDF model}
For the Gaussian PDF model, we assume that the PDF of the intrinsic magnetic field strength for cloud elements with the same volume densities ($n$) is a Gaussian function:
\begin{equation}\label{eq:bmodel_1}
P_{int}(B,\theta)=G(B,B_{mean},\sigma_{B_{int}}),
\end{equation}
where $\sigma_{B_{int}}$ is the dispersion of the intrinsic magnetic field strength, and $B_{mean}$ is the Gaussian mean magnetic field strength.
Combing \Crefrange{eq:pdf}{eq:bmodel_1}, the likelihood function for a single detection becomes a truncated Gaussian function:
\begin{equation}\label{eq:baye_conv_ga}
\begin{aligned}
P(B_{pos,i}|\theta)& =
  \begin{cases}
	\frac{G(B_{pos,i},B_{mean},\sigma_{con})}{\sigma_{con}(1-\Phi(B_{min},B_{mean},\sigma_{con}))} & B_{min} < B_{pos,i} \\
    0 &  B_{pos,i} < B_{min},  \\
  \end{cases}\\
\end{aligned}
\end{equation}
where $\Phi$ denotes the Gaussian cumulative distribution function, and $B_{min}= 0.5~\sqrt[]{4\pi \rho}\cdot0.2 (km/s)/25\degr$ represents our sample selection criteria. The quantity $\sigma_{con}^2 = \sigma_{B_{int}}^2 + \sigma_{B_{pos,i}}^2$ is the Gaussian sigma propagated from the dispersion of the intrinsic magnetic field strengths and the uncertainties in $B_{pos,i}$. Since $\sigma_{B_{pos,i}}$ is distinct for each $B_{pos,i}$ estimate, the width of the convolution $\sigma_{con}$ is not a constant. 

To determine how $B_{mean}$ varies with $n$, we tested two models.
Our first model, hereafter Model 1a, is composed of a power-law relation in high-density regions, and a constant field strength in low-density regions:
\begin{equation}\label{eq:bmodel_1a}
\begin{aligned}
 B_{mean}(n)& =
  \begin{cases}
    B_{0} & n < n_0 \\
	B_0(n/n_0)^{\alpha} & n > n_0,\\
  \end{cases}\\
\end{aligned}
\end{equation}
where $B_0$, $n_0$, and $\alpha$ are free model parameters. This model is suggested by \citet{cr10} based on their studies of numerous H\Rom{1} regions and molecular clouds.

Our second model, hereafter Model 1b, is a single power-law:
\begin{equation}\label{eq:bmodel_1b}
 B_{mean}(n)=B_0(n/150~\mathrm{cm^{-3}})^{\alpha},
\end{equation}
where $150~\mathrm{cm^{-3}}$ is a scale factor chosen so that the derived $B_0$ could be easily compared to other works. This model is suggested by recent observations that revealed the $B_{pos}$--$n$ relation within single molecular clouds \citep[e.g.,][]{ma12,ho17}.

\begin{figure*}
\includegraphics[width=\textwidth]{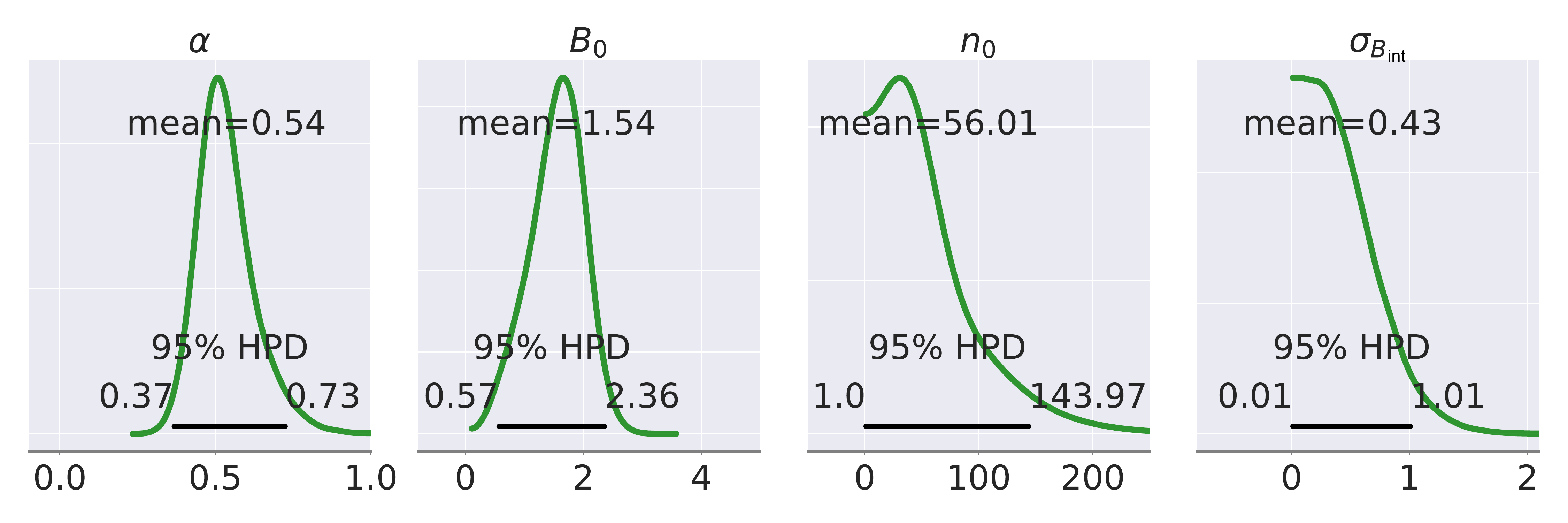}
\caption{Posteriors PDFs of each of the four parameters in the Bayesian analysis for Model 1a. The mean and 95\% highest posterior density (HPD) interval are labeled for each parameter.}\label{fig:baye_model_1a}
\end{figure*}
\begin{figure*}
\includegraphics[width=\textwidth]{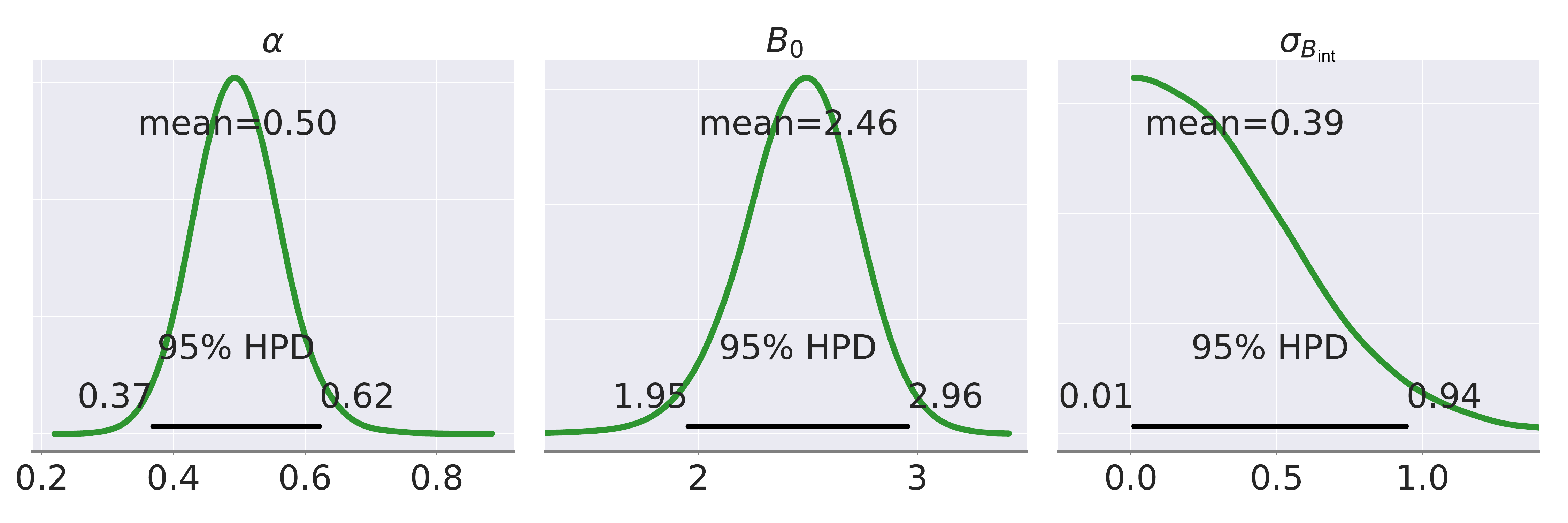}
\caption{Posteriors PDFs of each of the three parameters in the Bayesian analysis for Model 1b.}\label{fig:baye_model_1b}
\end{figure*}

To perform the Bayesian statistics, we choose priors with uniform PDFs across reasonable ranges:
\begin{equation}
\begin{aligned}
  \begin{cases}
 P(\alpha)& = U(\alpha,0.1,1)\\
 P(n_0)& = U(n_0,1,1000)\\
 P(B_0)& = U(B_0,1,100)\\
 P(\sigma_{B_{int}})& = U(\sigma_{B_{int}},0.01,1),
  \end{cases}
\end{aligned}
\end{equation}
where U(x,lower,upper) denotes a uniform probability function:
\begin{equation}\label{eq:uPDF}
\begin{aligned}
	U(x,lower,upper)=&
  \begin{cases}
    \frac{1}{upper-lower}, & lower < x < upper \\
	0 & otherwise.\\
  \end{cases}\\
\end{aligned}
\end{equation}
We note that the choices of these priors have only little effect on the results, because we provide sufficient data to constrain the model. 

With the likelihood function, constructed from \autoref{eq:pdf} to \autoref{eq:bmodel_1b}, and the priors described above, we used the Markov Chain Monte Carlo (MCMC) method to explore the free parameter space, and found the relative probabilities of all parameter sets. This MCMC method was performed using the Python PyMC3 Package \citep{sal16} with the No-U-Turn sample algorithm \citep{ho14}. We generated 40,000 total samples to explore the parameter space, and the first 10,000 samples were removed to allow burn-in. 

The posterior PDF of the model parameters for Model 1a and 1b are shown in \autoref{fig:baye_model_1a} and \autoref{fig:baye_model_1b}. 
The most probable values of $\alpha$ for Model 1a is 0.54 with a 95\% highest posterior density (HPD) interval from 0.37 to 0.73. The most preferred $\alpha$ for Model 1b is 0.50 with a 95\% HPD interval from 0.37 to 0.62. Both model $\alpha$'s are essentially identical at $\sim0.5$.

The posteriors of $\sigma_{B_{int}}$ in both models have peaks at $\sim$0, which indicates that the observed dispersion of $B_{pos}$ is likely lower than the observational uncertainties ($\sigma_{B_{pos}}$), and implies that the $B_{pos}$ uncertainties are possibly overestimated. Since the $B_{pos}$ uncertainties are dominated by the velocity dispersion uncertainty, calculated from the velocity dispersion measurements toward 11 filaments in \citet{ar13}, we speculate that velocity dispersion uncertainty is probably affected by the limited measurements or sample selection.

\subsubsection{Uniform PDF}\label{sec:uni_pdf}
For the Uniform PDF model, we assumed that the PDF of the intrinsic magnetic field strength for cloud elements with the same volume densities ($n$) is a uniform function. The probability function for observing $B_{pos}$ (\autoref{eq:pdf}) becomes a convolution of a uniform and a Gaussian distribution:
\begin{equation}\label{eq:baye_ga}
\begin{aligned}
P(B_{pos}|\theta)& =
  \begin{cases}
	\int{\frac{G(B_{pos},B,\sigma_{obs})}{B_{max}-B_{min}}}dB &  B_{min}< B_{pos} < B_{max} \\
    0 & otherwise. \\
  \end{cases}\\
\end{aligned}
\end{equation}
where $B_{max}$ represents the maximum magnetic field strength.

Similar to $B_{mean}$ in the Gaussian PDF models, we assume that $B_{max}$ scales with $n$ either by a broken power-law, hereafter Model 2a:
\begin{equation}\label{eq:bmodel_2a}
\begin{aligned}
 B_{max}(n)& =
  \begin{cases}
    B_{0} & n < n_0 \\
	B_0(n/n_0)^{\alpha} & n > n_0,\\
  \end{cases}\\
\end{aligned}
\end{equation}
or single power-law, hereafter Model 2b:
\begin{equation}\label{eq:bmodel_2b}
 B_{max}(n)=B_0(n/150~\mathrm{cm^{-3}})^{\alpha}.
\end{equation}
We used the same priors as in the Gaussian PDF models:
\begin{equation}
\begin{aligned}
  \begin{cases}
 P(\alpha)& = U(\alpha,0.1,1)\\
 P(n_0)& = U(n_0,1,1000)\\
 P(B_0)& = U(B_0,1,100),
  \end{cases}
\end{aligned}
\end{equation}

\begin{figure*}[!hbt]
\includegraphics[width=\textwidth]{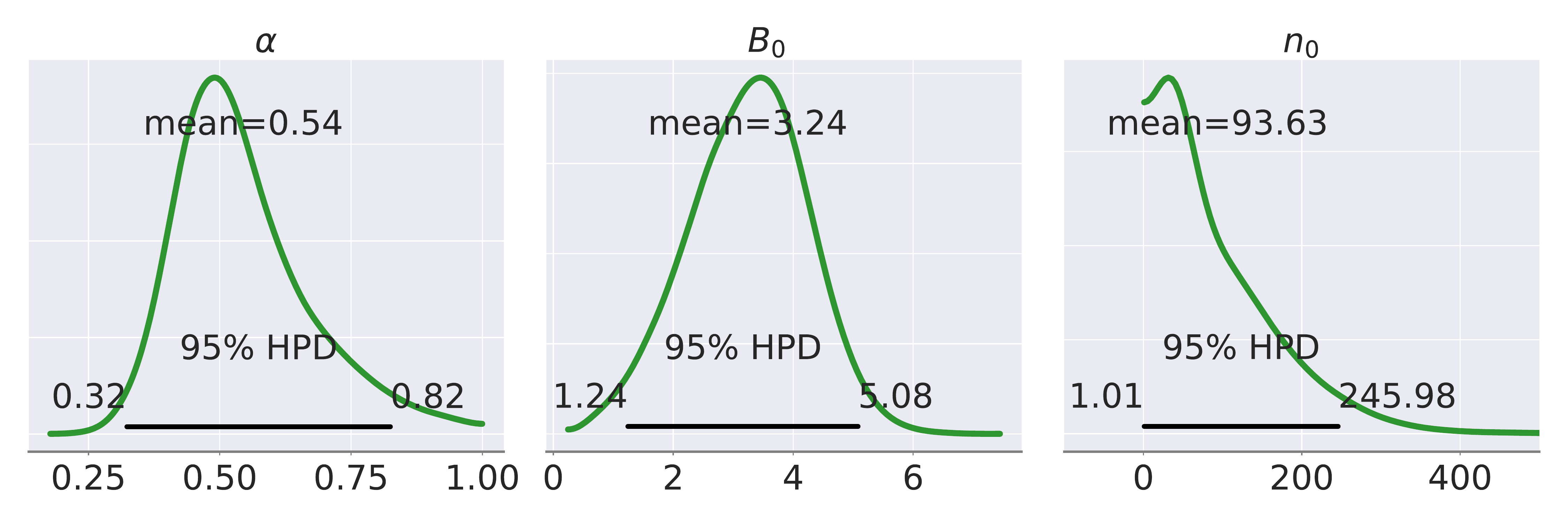}
\caption{Posteriors PDFs of each of the three parameters in the Bayesian analysis for Model 2a.}\label{fig:baye_model_2a}
\end{figure*}

\begin{figure}
\includegraphics[width=\columnwidth]{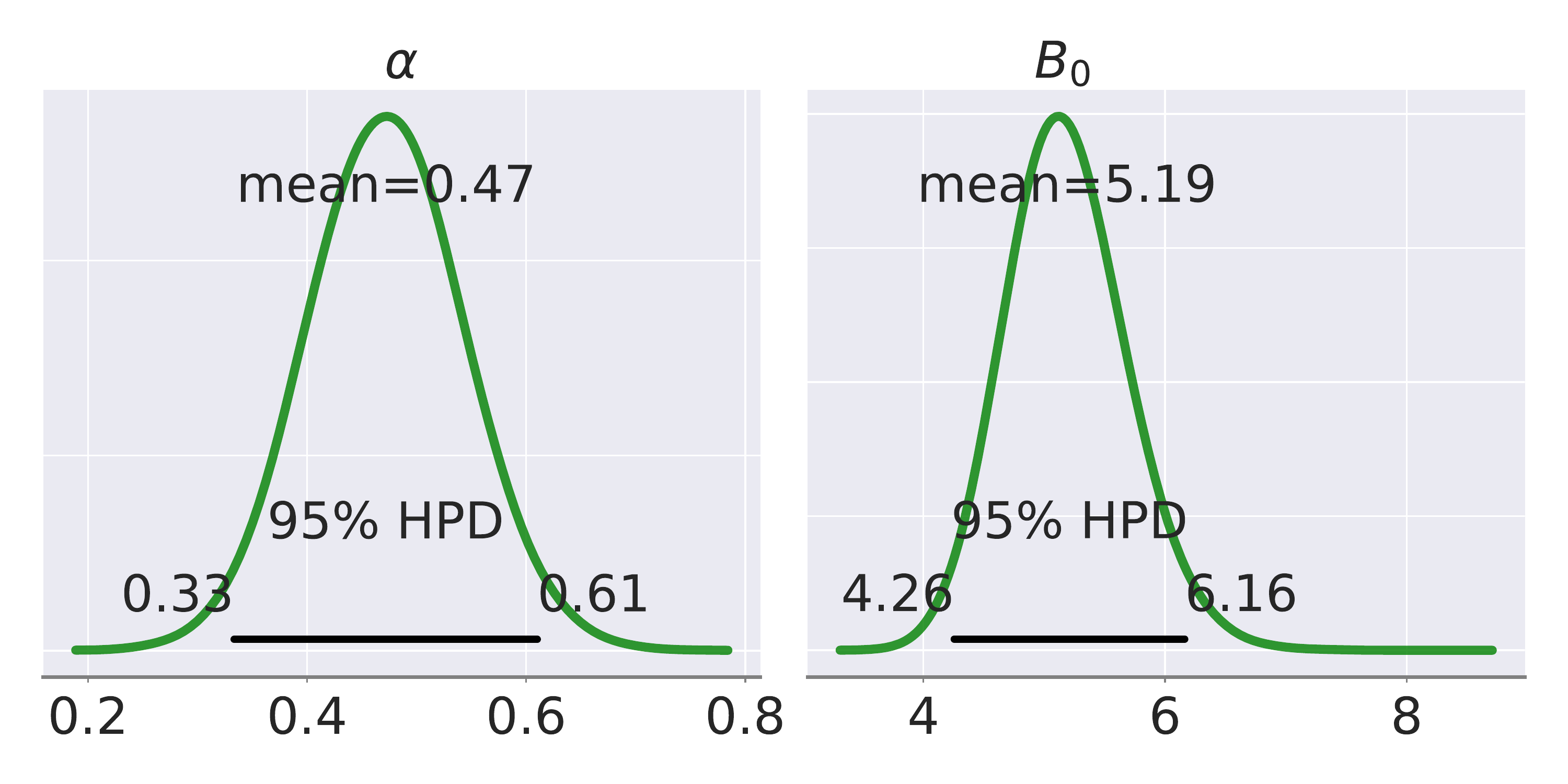}
\caption{Posteriors PDFs of each of the two parameters in the Bayesian analysis for Model 2b.}\label{fig:baye_model_2b}
\end{figure}

\begin{figure*}
\includegraphics[width=\textwidth]{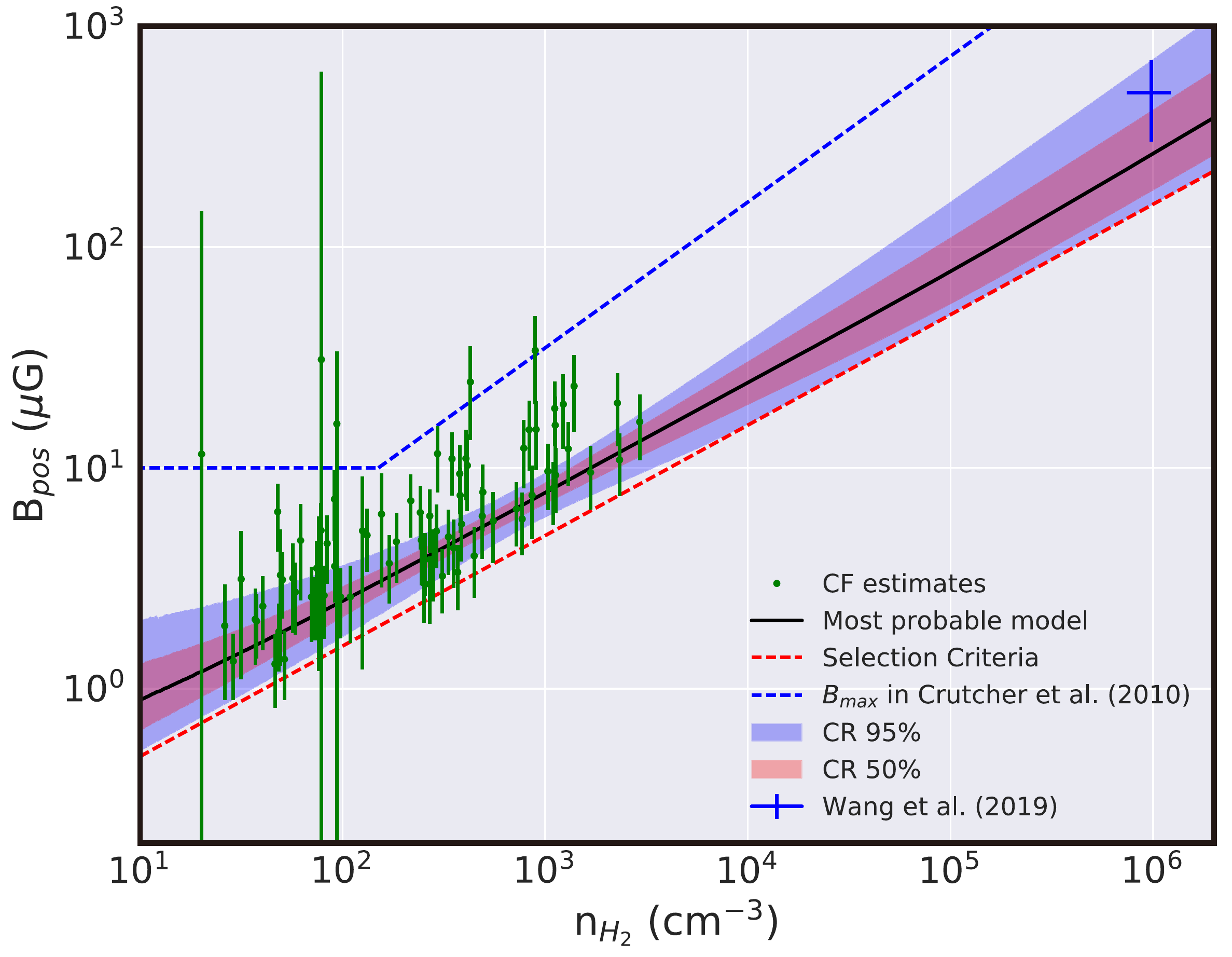}
\caption{B$_{pos}$ versus volume density with the posterior predictive distribution for Model 1b. The green points are the $\mathrm{R_c}$- and H-band magnetic field strength estimates. The blue and magenta regions mark the 50\% and 95\% confidence regions (CR), predicted by the posterior for Model 1b. The blue dashed line shows the $B_{max}$--$n$ relation obtained in \citet{cr10}. The blue cross at top right corner shows the B$_{pos}$ and $n$ obtained using the DCF method toward one of the dense hub-filament systems in the IC5146 cloud based on JCMT 850~$\mu$m thermal dust emission polarization data \citep{wa19}. This point matches our posterior predictive distribution.}\label{fig:Bvsn_gaussian}
\end{figure*}

We used the MCMC method to explore the parameter space for Model 2a and 2b. The posterior PDFs referred to the model parameters for Model 2a and 2b are shown in \autoref{fig:baye_model_2a} and \autoref{fig:baye_model_2b}. The most probable values of $\alpha$ for Model 2a and 2b are both $\sim$0.5, although the $\alpha$ for Model 2a has higher uncertainty. The posterior of $n_0$ for Model 2a shows a wide 95\% HPD interval ranging from 1 to 246 cm$^{-3}$ with somewhat at a peak at 94 cm$^{-3}$, which is more or less consistent with the $n_{H_2}=150$ cm$^{-3}$ in \citet{cr10}.

\subsubsection{Model Comparison}\label{sec:m_compare}
In order to evaluate the relative goodness of the tested models, we used the Watanabe-Akaike information criterion (WAIC, \citealt{wat10}) to rank these models. WAIC is a generalized version of the commonly used Akaike information criterion (AIC, \citealt{ak74}), and has been recently introduced for applications to astrophysics \citep[e.g.,][]{ra16,vi17}. While AIC only uses the maximum likelihood and number of model parameters to evaluate the quality of a model, WAIC ranks models using the averaged likelihood weighted by the posterior distribution and the effective degrees of freedom, estimated from the posterior variance. Hence, WAIC has been proposed as a fully Bayesian approach of estimating the predictive accuracy. Similar to AIC, the lower the WAIC value, the more predictive power the model has. 

The calculated WAIC values and the uncertainties in the WAIC difference ($\sigma_{\Delta \mathrm{WAIC}}$) for all the tested models are shown in \autoref{tab:model}. The Model 1b better explains our data, based on its lowest WAIC value. The WAIC values of the models assuming Gaussian PDF of $B$ (Model 1a and 1b) are significantly lower than the WAIC values of the Uniform PDF models (Model 2a and 2b), indicating that the assumption of Gaussian distributed $B$ is better than a uniform distributed $B$. 

Although the Gaussian distributed $B$ can better explain our data, the 95\% HPD interval of $\alpha$ of 0.33--0.61 for Model 2b is, nevertheless, similar to 0.37--0.62 range for Model 1b. The consistent $\alpha$ derived from $B_{max}$ and $B_{mean}$ indicate that both the mean and the upper boundary of the magnetic field strength scale with volume density by similar slopes, which disfavors the possibility that multiple trends are present. 

\autoref{fig:Bvsn_gaussian} shows the comparison between our data and the intrinsic $B$--$n$ distribution predicted by the posterior for Model 1b. We note that the observational uncertainties are not included in the plotted prediction, since the observational uncertainties vary by data point. The 95\% confidence regions seem to be consistent with the location of our data, to within their uncertainties. 

To test the prediction accuracy of our model, we added a point based on the volume density and magnetic field strength values estimated using the DCF method applied to JCMT 850~$\mu$m dust emission polarization data \citep{wa19}, toward a dense hub-filament system embedded in the IC5146 cloud, which was not included in the previous Bayesian analysis. Their polarization efficiency analysis shows that the 850~$\mu$m dust emission polarization data trace the aligned dust within the very dense clumps with $A_V$ up to 300 mag, and still their estimate is consistent with the prediction of Model 1b.

\begin{deluxetable*}{cccccccccc}
\tablecaption{Model Comparison\label{tab:model}}
\renewcommand{\thetable}{\arabic{table}}
\tablenum{1}
\tablehead{\colhead{Model Name} & \colhead{PDF of $B$} & \colhead{$B$--$n$ Relation} & \colhead{$\alpha$\tablenotemark{a}} & \colhead{$B_0$\tablenotemark{a}} & \colhead{$n_0$\tablenotemark{a}} & \colhead{$\sigma_{B_{int}}$\tablenotemark{a}} & \colhead{WAIC\tablenotemark{b}} & \colhead{$\Delta \mathrm{WAIC}$\tablenotemark{c}} & \colhead{$\sigma_{\Delta\mathrm{WAIC}}$ \tablenotemark{d}}
}
\startdata
Model 1a & Gaussian & Power-law + Flat  & $0.54\substack{+0.19 \\ -0.17}$ & $1.54\substack{+0.82 \\ -0.97}$ & $56\substack{+88 \\ -56}$ & $0.43\substack{+0.58 \\ -0.43}$ & 325.0 & 2.6 & 1.25 \\
Model 1b & Gaussian & Power-law  & $0.50\substack{+0.12 \\ -0.13}$ & $2.46\substack{+0.50 \\ -0.51}$ & ... & $0.39\substack{+0.55 \\ -0.39}$ & 322.4 & 0 & 0 \\
Model 2a & Uniform & Power-law + Flat  & $0.54\substack{+0.28 \\ -0.22}$ & $3.24\substack{+1.84 \\ -2.00}$ & $94\substack{+152 \\ -94}$ & ... & 362.8 & 40.5 & 3.3 \\
Model 2b & Uniform & Power-law   & $0.47\substack{+0.14 \\ -0.14}$ & $5.19\substack{+0.97 \\ -0.93}$ & ... & ... & 364.9 & 42.6 & 3.0 \\
\enddata
\tablenotetext{a}{The most preferred values. The uncertainties denote half of the 95\% HPD interval.}
\tablenotetext{b}{The WAIC value of each model. A lower WAIC indicates a better model.}
\tablenotetext{c}{Difference in WAIC relative to Model 1b.}
\tablenotetext{d}{Uncertainties in $\Delta$WAIC.}
{\addtocounter{table}{-1}}
\end{deluxetable*}

\subsubsection{Mass-to-Magnetic Flux ratio}\label{sec:m2f}
The mass-to-flux ratio ($M/\Phi$) of a region indicates the relative importance between the magnetic field and gravity. The $M/\Phi$ is defined as
\begin{equation}
(M/\Phi)=\frac{\mu m_{H}N(H_2)}{B},
\end{equation}
 \citep{mo76} where $\mu$=2.8 is the mean molecular weight per H$_2$ molecule. The mass-to-flux ratio is commonly compared to the critical mass-to-flux value to create a mass-to-flux ratio criticality ($\lambda$):
\begin{equation}
\lambda = \frac{(M/\Phi)}{(M/\Phi)_{cri}}.
\end{equation}
where $(M/\Phi)_{cri}$ is the critical mass-to-flux ratio
\begin{equation}
(M/\Phi)_{cri}=\frac{1}{2\pi \sqrt{G}}
\end{equation}
\citep{na78}.

Because the $M$ and $\Phi$ in these equations refer to the total magnetic field strength and column density along the magnetic field direction, the observed mass-to-flux ratio ($(M/\Phi)_{obs}$) is expected to be overestimated by a factor of $f$ due to projection effects:
\begin{equation}
(M/\Phi)=f (M/\Phi)_{obs}.
\end{equation}
The factor $f$ depends on the geometry of the clump and magnetic fields. Due to the unknown inclination angle of magnetic fields, \citet{cr04} suggested that a statistically average factor of $f=\frac{1}{3}$ could be used to estimate the real mass-to-flux ratio, accounting for the random inclinations for an oblate spheroid core, flattened perpendicular to the orientation of the magnetic field. In comparison, \citet{pl16} suggest a $f$ of $\frac{3}{4}$ for a prolate spheroid elongated along the orientation of the field.
Since we found that the IC5146 main filament is perpendicular to the mean magnetic field orientation, we adopt a factor of $\frac{1}{3}$ to correct the mass-to-flux ratio.

\begin{figure}
\includegraphics[width=\columnwidth]{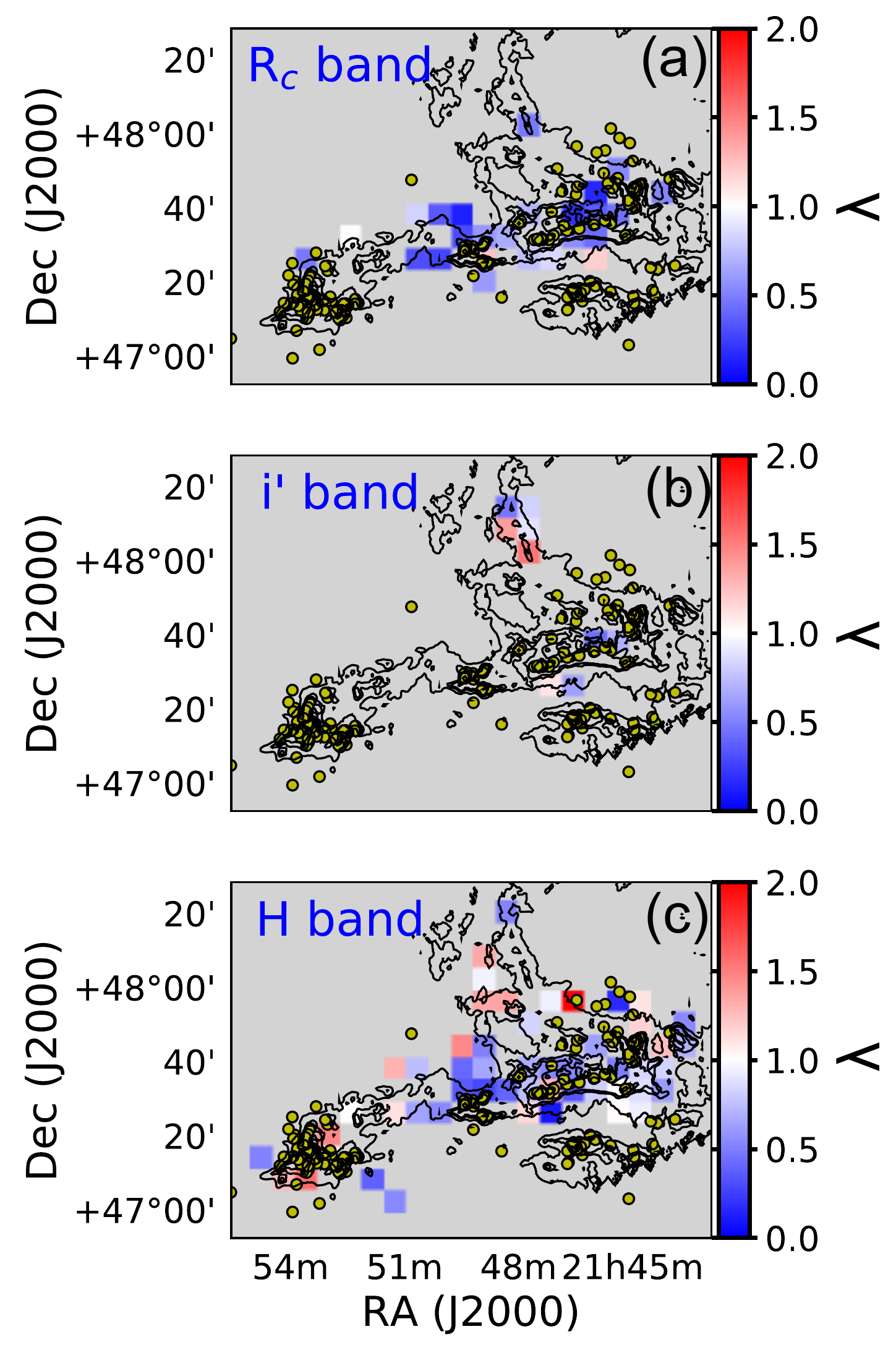}
\caption{Mass-to-flux criticality ratio maps over the IC5146 cloud for the $\mathrm{R_c}$-, $\mathrm{i'}$-, and H-band data. These mass-to-flux criticality ratios have been multiplied by a factor of $1/3$ to correct for average projection effects. The $\mathrm{R_c}$- and $\mathrm{i'}$-band data, tracing the magnetic fields in the outer part of the cloud, seem to show lower mass-to-flux ratios than do the H-band data. The yellow points identify locations of the YSO candidates found by \citet{ha08}. The black contours are the \textit{Herschel} 250~$\mu$m intensities with levels of 0.1, 1, and 5 Jy/beam.}\label{fig:m2fmap}
\end{figure}

We plotted the corrected mass-to-flux criticality ($\lambda$) derived from the $\mathrm{R_c}$-, $\mathrm{i'}$-, and H-band data in \autoref{fig:m2fmap}. The median uncertainty of the mass-to-flux criticality is $36\%$, which is similar to the uncertainties in $B_{pos}$ because the uncertainties in column densities, averaged to $3\times3$ arcmin pixels, are relatively minor. In the $\mathrm{R_c}$ and $\mathrm{i'}$ maps, most of the regions are magnetically subcritical ($\lambda<1$), suggesting that the magnetic field is sufficient to balance the gravity. In contrast, the H-band mass-to-flux criticality map reveals that some of the filaments are transcritical ($\lambda\approx 1$) or even supercritical ($\lambda>1$), and those regions are consistent with the spatial distribution of the YSOs identified in \citet{ha08}.

\begin{figure}
\includegraphics[width=\columnwidth]{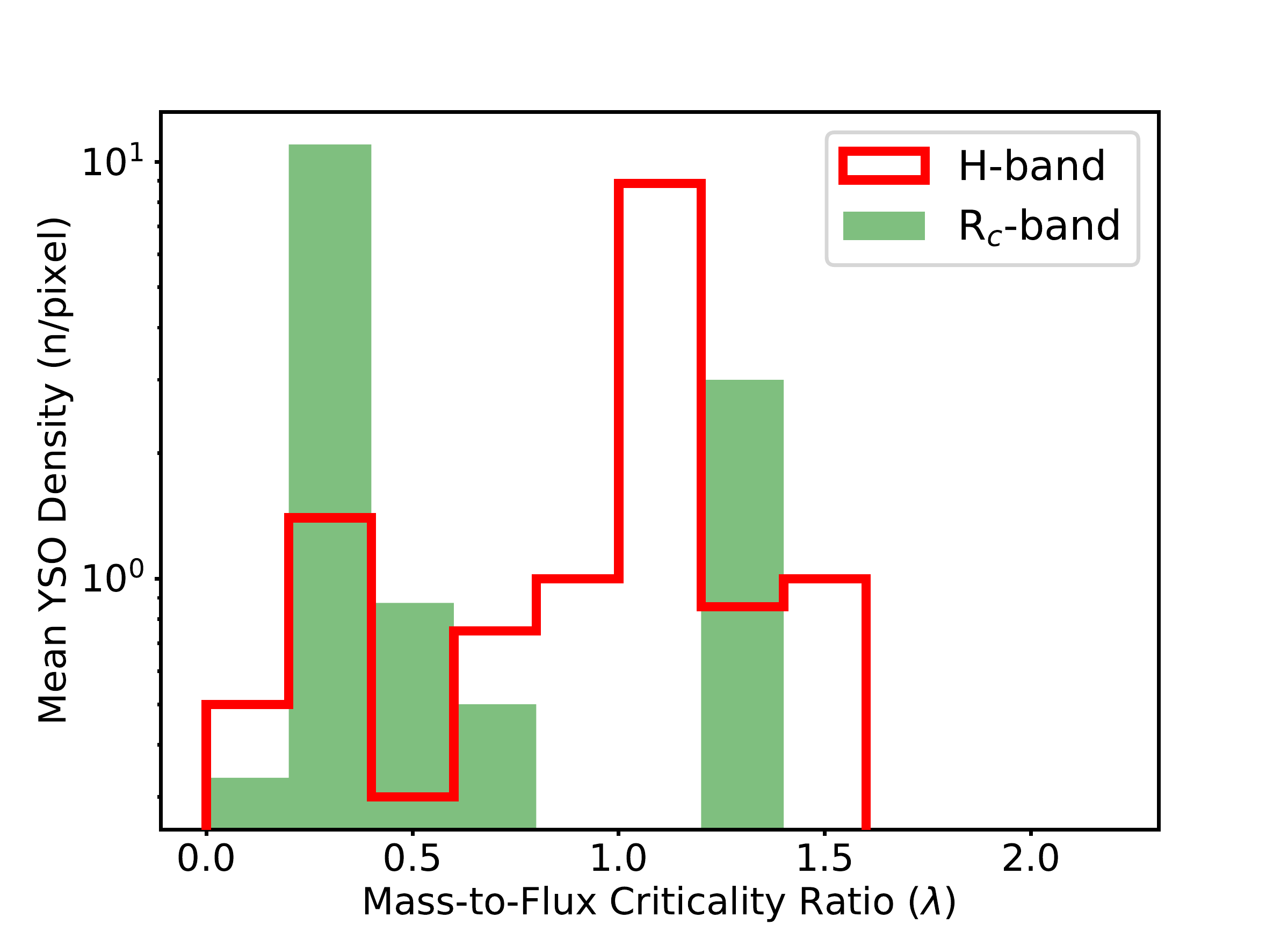}
\caption{Histogram of mean YSO densities with given measured mass-to-flux ratios. The y-axis denotes the log of the mean YSO numbers per pixel in regions having mass-to-flux ratios in 0.2 $\lambda$ bins. The green and red histograms represent the mass-to-flux ratios estimated using the $\mathrm{R_c}$- and H-band binned data, respectively. Most YSOs tend to be located near or within magnetically transcritical regions ($0.7<M/\Phi<1.5$), traced by the H-band data. The mass-to-flux ratios estimated from the $\mathrm{R_c}$-band are mostly subcritical, showing a clear change in the distributions, with the peak moving to larger mass-to-flux ratios from $\mathrm{R_c}$- to H-band.
}\label{fig:YSO_m2f}
\end{figure}

In order to investigate how mass-to-flux criticality connects to star formation, \autoref{fig:YSO_m2f} shows the histogram of the mean YSO number density with given mass-to-flux criticality estimated in the $\mathrm{R_c}$- and H-bands. These YSO densities were averaged from each $6\times6$ arcmin pixel within given mass-to-flux criticality range, and indicate the relative probabilities to form stars per unit area. The size of the $\lambda$ histogram bins is 0.2, which is comparable to our uncertainties in mass-to-flux criticality of $\sim$0.2--0.4. The $\mathrm{i'}$-band results are not included, because the pixels with $\mathrm{i'}$-band mass-to-flux estimates cover only a small area. The greatest YSO densities are located in the H-band pixels with a mass-to-flux criticality of 0.6--1.6, with a peak at 1.0--1.2. The YSO density peak near mass-to-flux criticality of 1.0-1.2 favors the scenario where star formation mainly takes place in $\lambda\approx 1$ regions. 

In comparison, the $\mathrm{R_c}$-band mass-to-flux criticalities spanned by high YSO densities are mostly subcritical. These clearly different $\mathrm{R_c}$- and H-band distributions show that the polarization data at different wavelengths trace different parts of the clouds: the H-band data tend to trace the supercritical regions, while the $\mathrm{R_c}$-band data trace the subcritical regions.

\begin{figure}
\includegraphics[width=\columnwidth]{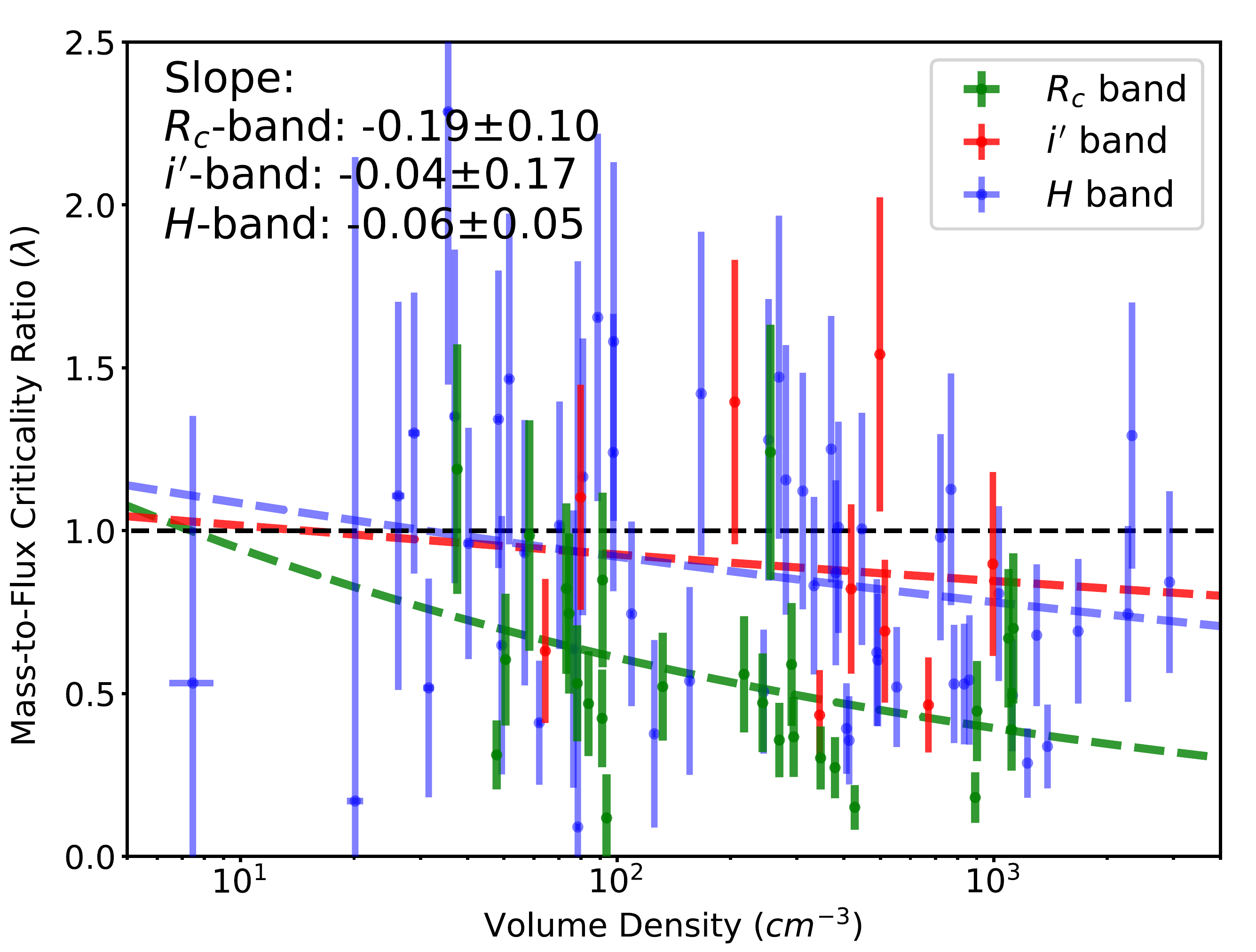}
\caption{Mass-to-flux criticality ratios versus volume densities for the $\mathrm{R_c}$-, $\mathrm{i'}$-, and H-band data. The H-band mass-to-flux ratios range from 0.1 to 1.5, while the $\mathrm{R_c}$- and $\mathrm{i'}$-band Mass-to-flux ratios tend to be subcritical, ranging from 0.1 to 1. The dashed lines show the best-fit power-laws for the three data sets. 
}\label{fig:m2vsden}
\end{figure}

To investigate how mass-to-flux criticality evolves within the clouds, we plot the mass-to-flux criticality versus volume density in \autoref{fig:m2vsden}. We calculated the slope of mass-to-flux criticality and volume density, for the estimates based on the $\mathrm{R_c}$-, $\mathrm{i'}$-, and H-band data. All the calculated slopes are close to zero, suggesting that the variation of mass-to-flux criticality is not strongly correlated with volume density.

\section{DISCUSSION}\label{sec:discussion}
\subsection{Magnetic Field Morphology}\label{sec:Bmor_dis}
The background starlight polarizations observed towards the IC5416 cloud reveal detailed information about the magnetic fields within and around the cloud. The spatially averaged polarization map further provides an evenly-sampled magnetic field morphology on an 0.6 parsec-scale. At this scale, the mean polarization angles over the cloud are very similar and the bin-to-bin angular dispersion is small ($<18\degr$), indicating that the overall parsec-scale magnetic field around this cloud is fairly uniform. In addition, we further find that the parsec-scale magnetic field is roughly perpendicular to the main filament, but the filament in the northern region is parallel to the mostly uniform magnetic field.

Numerical simulations of filament formation often show alignment between filaments and magnetic fields via different mechanisms. \citet{na08} modeled the evolution of a magnetically subcritical cloud, and found a sheet-like or filamentary cloud formed via gas condensation along the large-scale magnetic field lines. The major clouds formed were perpendicular to the large-scale magnetic fields, and some diffuse filaments (striations) might still flow along, and thus parallel, to the magnetic fields. \citet{va14} simulated the formation of filaments from a self-gravitating layer threaded by magnetic fields in both subcritical and supercritical conditions The forming filaments appeared to be either a network of hubs and converging filaments (for weak magnetic fields) or a network of parallel filaments perpendicular to the magnetic field (for strong magnetic fields). On the other hand, simulations assuming weak magnetic fields can also generate filaments parallel to magnetic fields, if the gas and magnetic fields are both shaped by turbulent compression \citep[e.g.,][]{pa02} or dragged by the gas flow under the action of gravity \citep[e.g.,][]{go18}. 

The uniformity of magnetic fields can be used to evaluate the relative importance of magnetic fields and turbulence. \citet{wa19} showed that the magnetic field angle dispersion ($\delta\phi$ in radians) is approximately proportional to the Alfv\'{e}nic Mach number ($M_A$), under the assumptions of the DCF method, by
\begin{equation}
M_A= \frac{\delta \phi \cdot \sin\theta}{Q},
\end{equation}
where $\theta$ denotes the inclination angle of the magnetic field. Adopting a Q of 0.5, the angular dispersion observed of $18\degr$ corresponds to an Alfv\'{e}nic Mach number of 0.6~$\sin\theta$ (i.e. sub-Alfv\'{e}nic for any $\theta$), indicating that the turbulence is relatively weaker than the magnetic fields in this cloud. Hence, we rule out strong turbulence (super-Alfv\'{e}nic) models. 

\begin{figure*}
\includegraphics[width=\textwidth]{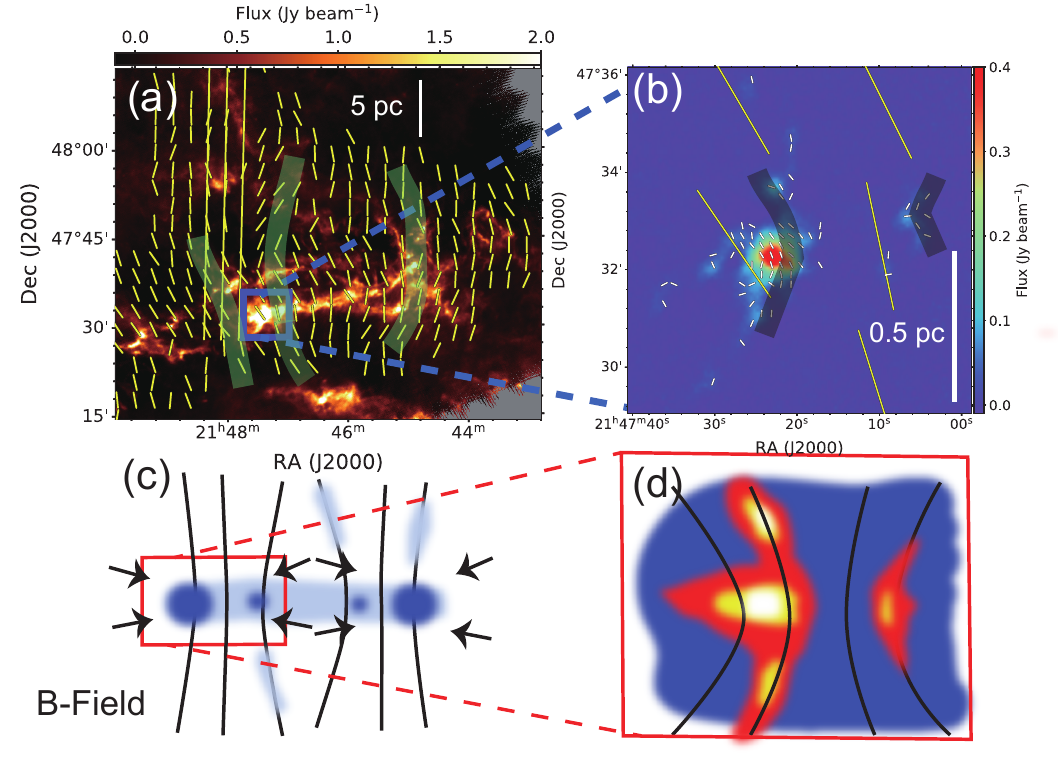}
\caption{Cartoon figures and observational counterparts illustrating the self-similar magnetic field morphologies around the IC5146 dark clouds at different scales. (a) The spatially averaged H-band polarization map overlaid on the $Herchel$ 250~$\mu$m image. The polarization map shows a curved magnetic field morphology at $\sim$5 pc scale toward the ends of the main filament, marked by the shadowed regions. (b) The JCMT POL-2 850~$\mu$m continuum polarization map \citep{wa19}. The 850~$\mu$m polarization (white segments, rotated by 90\degr to infer magnetic field orientations) shows a sub-parsec scale hourglass morphology (shadowed regions) toward the center of the two clumps. (c) The pc-scale hourglass magnetic field morphology dragged by the contraction of the main filament toward the two ends. (d) The sub-parsec scale magnetic field dragged toward the center between the two clumps.
}\label{fig:cartoon}
\end{figure*}

In addition, $Herschel$ studies found that the IC5146 parsec-scale main filament consists of multiple parallel filaments perpendicular to the parsec-scale magnetic field \citep{ar11}. This geometry is consistent with the predictions of the simulation of a subcritical, self-gravitating cloud \citep{va14}. However, some hub-filament structures have been found embedded within the parsec-scale filament. \citet{wa19} observed one of the sub-parsec scale hub-filament systems located at the end of the main filament with the JCMT, and found a curved magnetic field, possibly dragged by contraction along the parsec-scale main filament, as shown in \autoref{fig:cartoon}(a) and (b). These results are consistent with the simulations of supercritical conditions \citep{va14}. With the above observed features, we speculate that the importance of magnetic fields in the IC5146 cloud is scale-dependent. The parsec-scale filaments possibly formed in magnetically subcritical condition, but these filaments gradually became supercritical at the sub-parsec scale, evolve into hub-filament systems, and form stars.

The curved magnetic field morphology found in the western part of the main filament supports this notion. \autoref{fig:cartoon}(a) and (b) shows the comparison between the pc-scale and the subparsec-scale polarization morphologies toward the western main filament. The pc-scale polarization PA near the filament ridge is perpendicularly aligned with the main filament, but gradually becomes misaligned when the distance from the filament ridge increases. This morphology is self-similar to the morphology of the hourglass magnetic field morphology at subparsec-scale, and hence the large-scale curvature is also possibly dragged by the contraction of the parsec-scale main filament. \autoref{fig:cartoon}(c) and (d) show cartoon figures illustrating the self-similar morphologies described above. Nevertheless, we cannot rule out the possibility that the curvature pattern is possibly tracing the magnetic field from two different layers of clouds, because, as shown in (\autoref{sec:dist}), the IC5146 cloud likely consists of two layers at distances of $\sim 600$ and 800 pc. 

\subsection{Magnetic Field Strength and Mass-to-flux Ratio}\label{sec:Bstr_dis}
Whether magnetic fields within molecular clouds are sufficiently strong to support the gas against gravity and prevent collapse is a key question in star formation. The theories assuming strong magnetic fields expect that a molecular cloud is initially magnetically-subcritical but gradually becomes supercritical as density increases through magnetic flux diffusion mechanisms, such as ambipolar diffusion \citep[e.g.,][]{mo91}, accretion \citep[e.g.,][]{he14}, turbulent diffusion \citep[e.g.,][]{ki02}, or reconnection diffusion \citep[e.g.,][]{sa10}. One observationally testable prediction of the strong magnetic field theories is that the mass-to-flux criticality must be subcritical in cloud envelopes, whereas in cores diffusion mechanisms should lead to transcritical or supercritical condition \citep[e.g.,][]{el07}. In contrast, the weak magnetic fields theories often assume or require supercritical mass-to-flux ratios over entire molecular clouds \citep{cr12}. Previous observations have shown a transition from sub- to super-critical: \citet{ko12} revealed a transition from a super- to a sub-critical state as a function of distance from the emission peak within the W51 e2 core. With a sample of 50 star-forming clumps, \citet{ko14} found the same transition while the average magnetic fields gradually evolve from aligned to misaligned with the dust emission gradient. \citet{ho17} found a critical density of n$_H= 300-670$ cm$^{-2}$ within the G28.23-00.19 cloud, in which the cloud mass-to-flux ratio evolves from sub- to super-critical.

With our polarization data, we estimated the magnetic field strengths and mass-to-flux ratios over the IC5146 cloud. The mass-to-flux ratio estimated using our $\mathrm{R_c}$-band data are all subcritical. In contrast, the mass-to-flux ratios estimated using our H-band data show both subcritical and supercritical mass-to-flux ratios. The polarization we measured is the average magnetic field morphology integrated along the line of sight weighted by the local densities and polarization efficiencies, and hence the depth the polarization data can trace is sensitive to how polarization efficiency decays with density. In paper \Rom{1}, we showed that the polarization efficiency of our $\mathrm{R_c}$-band data, mostly taken from stars with $A_V<5$ mag, drops quickly as $A_V$ increases, with a power-law index of $-0.7$, suggesting that the $\mathrm{R_c}$-band polarization is mainly contributed from the cloud envelope ($A_V<<5$ mag). On the other hand, the polarization efficiency of our H-band data, mostly taken from stars with $A_V<15$ mag, drops smoothly as $A_V$ increase, with a power-law index of $-0.3$, indicating that with this data set it is possible to trace the magnetic fields into the cloud center. This conclusion is consistent with the different distributions of mass-to-flux ratios of YSOs estimated by the $\mathrm{R_c}$- and H-band data, shown in \autoref{fig:YSO_m2f}. As a result, we speculate that the IC5146 cloud envelope is subcritical, and some of the central regions could either remain subcritical or become supercritical, and this is where the YSOs tend to be located. These results favor the strong magnetic field star formation models. 

A remaining question is how the mass-to-flux ratio is caused to increase in this cloud. The classical ambipolar diffusion mechanism predicts an increase of the mass-to-flux ratio is with density from subcritical to supercritical \citep{mo99}, although this mechanism is more efficient in high density regions at small-scale \citep{he14}. The magnetic field regulated accretion mechanism also predicts an increase of the mass-to-flux ratio with density \citep{he14}. This is seen in some observations \citep[e.g.,][]{ho17}. However, here we do not find a significant correlation between the mass-to-flux criticality and density. Reconnection diffusion theory might explain the observed trend, because reconnection diffusion can also be efficient even in parsec-scale low-density regions and thus create supercritical diffuse clouds \citep{la12}. 

The lack of a correlation between the mass-to-flux criticality and the density could also be caused by the separation of the foreground or background clouds from the IC5146 dark streams. The difference in magnetic field geometry between separate clouds could contribute an additional polarization angle dispersion $\delta \Phi$, and thus the magnetic field strength estimated using the DCF method would be underestimated. This bias could be significant in the diffuse area, where the polarized intensities from the two cloud layers are comparable. This underestimation of magnetic field strength in low density regions could possibly cause the apparent relation between the mass-to-flux criticality and density to be flat or even negative.

\subsection{$B$--$n$ Relation}\label{sec:dis_Bvsn}
How magnetic field strength scales with cloud density provides important information on how cloud collapse may be regulated by magnetic fields. Based on the Zeeman effect measurements toward more than hundred clouds and cores, \citet{cr10} found that magnetic field strength scales with density by $B \propto n^{0.65}$ for $n_{H_2}>150$ cm$^{-3}$, and suggested that core contraction is isotropic and not likely regulated by magnetic fields. In addition, a constant magnetic field strength was found for $n_{H_2}<150$ cm$^{-3}$, which favored the scenario that gas accretion is guided by magnetic fields and increases the mass-to-flux ratio of clouds \citep{he14}. A known problem of the Zeeman analysis is that these measurements were obtained from different types of clouds that may not be in the same evolutionary sequence \citep{li15}. The OH Zeeman measurements were obtained from dark clouds while the CN Zeeman data were mostly from massive cluster-forming regions in giant molecular clouds; however, most dense cores in nearby dark clouds are not likely to evolve into massive cluster-forming clumps \citep{li14}. Hence, recent observations aim at revealing how the magnetic field strength varies with the density using data for single clouds. A diversity of $B$--$n$ relations is shown by those observations that probe magnetic field strength structures within single clouds. For example, \citet{ma12} and \citet{ho17} found a steep $B$--$n$ indices of 0.75$\pm$0.02 and 0.73$\pm$0.06 toward GRSMC 45.60+0.30 and IRDC G028.23-00.19, respectively. In contrast, \citet{li15} found a shallow index of 0.41$\pm$0.04 toward NGC 6334.

In \autoref{sec:Bvsn}, our Bayesian analysis favored $B \propto n^{0.50}$ with a 95\% HPD interval from 0.37 to 0.62. This 0.5 index is significantly lower than 2/3 (isotropic cloud collapse, \citealt{cr10}) and implies that cloud collapse tends to follow the magnetic field. This is consistent with a scenario where magnetic fields are important for impeding cloud collapse \citep[e.g.,][]{fi93,mo91,mo17}. This result supports our interpretation that magnetic fields are important in regulating the formation of the parsec-scale filaments, based on the alignment between magnetic field and filaments, and also on the mass-to-flux criticality estimate. We note that the IC5146 dark cloud and the Cocoon Nebula are likely at different distances, and so the IC5146 system is not a single-cloud system. However, the magnetic fields found in the darks cloud and the Cocoon Nebula have similar morphology, and thus we assume that the role of magnetic fields in these two regions are similar. This assumption is supported by the similar $B$--$n$ indices obtained in both Gaussian and Uniform PDF models.

Compared to the indices of 0.75$\pm$0.02 and 0.73$\pm$0.06 toward GRSMC 45.60+0.30 \citep{ma12} and IRDC G028.23-00.19 \citep{ho17}, we find a significantly shallower index, even though these other two studies also used near-infrared starlight polarization data, and probed parsec-scale clouds with a similar density range, 10--10${^2}$ cm$^{-3}$ for GRSMC 45.60+0.30 and 10${^2}$--10${^3}$ cm$^{-3}$ for G028.23-00.19. On the other hand, the index of 0.41$\pm$0.04 found by \citet{li15} in NGC 6334 is similar to our results, but their data trace the environments in denser regions (10$^3$--10$^7$ cm$^{-3}$) and on larger scales (10--0.1 pc). This diversity of $B$--$n$ relations possibly implies that filaments can form in both strong and weak magnetic field environments.

A remaining question is, ``which physical cloud properties determine the filament formation environment?''. Cloud masses are probably not the dominant factor. The most massive clump in GRSMC 45.60+0.30 and IRDC G028.23-00.19 have masses of 900 M$_\sun$ \citep{ma12} and 1520 M$_\sun$ \citep{sa13}, respectively, which are comparable to the mass of the Cocoon Nebula star-forming cluster in the IC5146 cloud (1309 M$_\sun$, \citealt{ha08}). NGC6334 hosts the most massive clump (2778 M$_\sun$, \citealt{ma08}) among all these four clouds; however, its $B$--$n$ slope favors the strong magnetic field condition. One different property between the weak and strong magnetic field case is that the GRSMC 45.60+0.30 and IRDC G028.23-00.19 clouds are both quiescent regions and possibly still young, while star formation is already ongoing in the IC5146 and NGC6334 clouds. One possible explanation is that the star formation feedback is correlated with magnetic fields, and causes anisotropic compression to surrounding gas. Nevertheless, recent numerical simulations of self-gravitating clouds show that the $B$--$n$ power-law index is likely insensitive to time once self-gravity has had time to act \citep{lp15}. Measurements of the $B$--$n$ relation toward more clouds are essential to provide a statistical basis to answer this question. 

If a diversity of $B$--$n$ relation is present among different clouds, the \citet{cr10} analysis might be biased by their uniform PDF model. Their index of 0.65 was estimated from the maximum magnetic field ($B_{max}$) and density ($n$), and $B_{max}$ is mainly determined by the upper envelope of the $B_{los}$--$n$ distribution. If a diversity of $B_{los}$--$n$ relations is present in their sample, then the upper envelope of the $B_{los}$--$n$ distribution would be dominated by only the steepest relation, in which $B_{los}$ is enhanced most efficiently. Hence, the obtained index is merely the upper-limit of all possible trends present in the sample set. This argument can also explain why a different index of 0.47 was found in \citet{cr99} using a subset of the samples used in \citet{cr10}. As an illustration, we compare our results in \autoref{fig:Bvsn_gaussian} with the $B_{max}$--$n$ upper envelope obtained in \citet{cr10}. This plot shows that most of our sample values are well below the $B_{max}$--$n$ upper envelope of \citet{cr10}. So while our results are actually consistent with the \citet{cr10} model, we found a very different slope.

\subsection{Caveats in the Magnetic Field Strength Estimation}\label{sec:caveat}
In addition to the reported observational uncertainties of $\sim35\%$ in the magnetic field strengths, the assumptions we made will also introduce some systematic uncertainties and bias. The major systematic uncertainties originate with the question of whether the polarization, the density, and the gas velocity dispersion data really probe the same space in the cloud, since they were taken using different observational methods.

First, to estimate the mean volume density along the line of sight, a boundary needs to be adopted for the cloud; however, clouds do not have physically well-defined  boundaries, and thus numerous methods  have been proposed to define artificial or effective boundaries \citep[e.g.,][]{cr04,ma12,ho17}. \citet{ho17} show that the choice of boundary can significantly change the estimated magnetic field strength and the $B$--$n$ slope. 

Ideally, we expect to set a boundary within which dust contributes to most of the observed polarization, and hence the calculated mean density can really match the same regions which are traced by our polarization data. To achieve this, knowledge of dust alignment efficiency along the line of sight is necessary, though modeling dust alignment in 3D space is challenging. \citet{os01} performed a simulation of evolving molecular clouds, and tested how the simulated polarization angle dispersion could match the density structure, in order to evaluate the validity of the DCF method. However, they assumed a constant polarization efficiency, which is inconsistent with modern radiative alignment torque theory \citep[e.g.,][]{la07,wh08}. \citet{se19} simulated synthetic dust polarization maps of molecular clouds based on the modern radiative alignment torque theory, and suggested that the observed polarization patterns resemble best the mass-weighted, line-of-sight averaged field structure, at least for observing wavelengths $>160~\mu m$. However, their simulation does not include optical or near-infrared wavelengths, for which the polarization efficiency is expected to drop in the dense regions \citep{la07,wh08}. Furthermore, they do not yet investigate how well the polarization angle dispersion represents the real complexity of magnetic fields. In this paper, we have assumed the measured polarization angle dispersion tends to trace the high-density regions, and we have defined an effective thickness that accounted for the central regions with the highest local volume densities; however, this assumption still needs to be tested.

Second, we do not have velocity information over the entire IC5146 cloud, and thus our estimation relies on the finding in \citet{ar13} that the gas velocity dispersion is roughly a constant for low-density filaments, based on the IRAM 30~m observations of $^{13}$CO (2-1), C$^{18}$O (2-1) and N$_2$H$^+$ (1-0) lines toward several of the filaments. Although some of their samples have low column densities ($\sim2\times10^{21}$ cm$^{-2}$), these molecular lines are high-density tracers due to their high critical densities, and hence whether the gas velocity dispersion is a constant in the diffuse regions might still be an open question. If the large-scale diffuse regions are more turbulent than the high-density regions, then according to the Larson's law our velocity dispersions, and the inferred magnetic field strengths, in the diffuse regions could be underestimated. Nevertheless, this implies a shallower  $B$--$n$ slope and a lower mass-to-flux criticality in the diffuse areas, both of which enforce our conclusion that the IC5146 molecular cloud forms in a strong magnetic field environment.

\section{SUMMARY}
\label{sec:summary}
\begin{enumerate}
\item Analyzing the new $Gaia$ DR2 data, we find that the IC5146 cloud likely consists of two separate clouds; the first cloud, including the densest main filament, is at a distance of $\sim600$ pc, and the second cloud, associated with the Cocoon Nebula, is at a distance of $\sim800$ pc.

\item The spatially averaged H-band polarization map shows an organized parsec-scale magnetic field over the entire IC5146 cloud. The magnetic field is perpendicular to the main filament structure, but parallel to the more diffuse filament extending to the north. The angular dispersion of the polarization position angle is generally higher in the western part ($\sim 18\degr$) than in the eastern part ($\sim 12\degr$), possibly caused by some large-scale gas motion or the multiple layers along the line of sight.

\item We estimate the magnetic field strength using the Davis-Chandrasekhar-Fermi method, and investigate how the magnetic field strength scales with volume density. We use a Bayesian approach to fit the $B_{pos}$--$n$ relation with either a single power-law or a broken power-law, using either a truncated Gaussian or a uniform likelihood function. We find that the single power-law model with a truncated Gaussian likelihood function can better explain the observed data. The derived power-law index ($\alpha$) is $\sim0.50$ with a 95\% highest posterior density interval from 0.37 to 0.62. This is significantly smaller than the value 0.66 which would indicate isotropic collapse. Our result suggests that collapse of the IC5146 cloud tends to follow, and thus is regulated, by the magnetic fields as predicted by strong magnetic field models.

\item We estimate the mass-to-flux ratio using the $\mathrm{R_c}$-, $\mathrm{i'}$-, and H-band data. Because of the wavelength-dependent decay rate of the polarization efficiency, the data in the optical bands tend to trace the magnetic field in the outer parts of the cloud, while the infrared data can trace deeper. The derived mass-to-flux ratios are generally subcritical in the cloud envelope, as traced by $\mathrm{R_c}$-band data. In contrast, some of the deep regions, traced by H-band data, become transcritical or supercritical. We further show that the identified YSOs in this cloud are mostly distributed in areas with mass-to-flux criticality of 0.6--1.5. This is consistent with a scenario where the cloud periphery is generally supported by magnetic fields, but gravity in the central regions eventually overcomes magnetic fields and triggers star formation.

\end{enumerate}

\acknowledgments
This research was conducted in part using the Mimir instrument, jointly developed at Boston University and Lowell Observatory and supported by NASA, NSF, and the W.M. Keck Foundation. This work and the analysis software for Mimir data were developed under NSF grants AST 06-07500, 09-07790, 14-12269, and 18-14531 to Boston University. J.W.W., S.P.L., and C.E. are thankful to the support from the Ministry of Science and Technology (MOST) of Taiwan through the grants MOST 105-2119-M-007-021-MY3, 105-2119-M-007-024, and 106-2119-M-007-021-MY3. J.W.W. is a University Consortium of ALMA--Taiwan (UCAT) graduate fellow supported by the Ministry of Science and Technology (MOST) of Taiwan through the grants MOST 105-2119-M-007-022-MY3. P. M. K. acknowledges support from MOST 108-2112-M-001-012 and MOST 107-2119-M-001-023, and from an Academia Sinica Career Development Award. E.C. acknowledges (i) The National Key R\&D Program of China grant no. 2017YFA0402600, (ii) The International Partnership Program of Chinese Academy of Sciences grant no. 114A11KYSB20160008, and (iii) partial support by Special Funding for Advanced Users, budgeted and administrated by Center for Astronomical Mega-Science, Chinese Academy of Sciences (CAMS).

\software{Aplpy \citep{ro12}, Astropy \citep{asp13}, NumPy \citep{va11}, PyMC3 \citep{sal16}, SciPy \citep{jo01}}

\end{document}